\documentclass[a4paper,fleqn,usenatbib,useAMS]{mnras}

\usepackage{graphicx}	% Including figure files
\usepackage{amsmath}	% Advanced maths commands
\usepackage{amssymb}	% Extra maths symbols
\usepackage{multicol}        % Multi-column entries in tables
\usepackage{bm}		% Bold maths symbols, including upright Greek
\usepackage{pdflscape}	% Landscape pages
\usepackage{threeparttable}
\newcommand\pname{K2-139\,b}
\newcommand\ssname{K2-139}
\newcommand\sname{EPIC\,218916923\,}
\newcommand{\kms}{\,km\,s$^{-1}$} % km per second
\newcommand{\ms}{\,m\,s$^{-1}$} % m per second
\newcommand{\smass}[1][$M_{\odot}$]{ $ 0.919 \pm 0.033 $ #1} 
\newcommand{\sradius}[1][$R_{\odot}$]{ $0.862 \pm 0.032 $ #1}

\newcommand{\Tzerob}[1][days]{$7325.81714 \pm 0.00033 $#1} 
\newcommand{\Pb}[1][days]{$28.38236 \pm { 0.00026 }$ #1} 
\newcommand{\esinb}[1][]{$0.10 _{ - 0.30 } ^ { + 0.29  }$ #1 } 
\newcommand{\ecosb}[1][]{$0.06 _{ - 0.27 } ^ { + 0.24  }$ #1 } 
\newcommand{\eb}[1][]{$0.12 _{ - 0.08 } ^ { + 0.12 }$ #1 }      
\newcommand{\wb}[1][deg]{$124 _{ - 79 } ^ { + 175 }$ #1} 
\newcommand{\bb}[1][]{$0.30 _{ - 0.19 } ^ { + 0.21 }$ #1}       
\newcommand{\arb}[1][]{$44.8 _{ - 6.7 } ^ { + 4.7 }$ #1}       
\newcommand{\rrb}[1][]{$0.0961 _{ - 0.0015 } ^ { + 0.0023  }$ #1}       
\newcommand{\kb}[1][$m \, s^{-1}$]{$27.7 _{ - 5.3  } ^ {+ 6.0 }$ #1} 
\newcommand{\ib}[1][deg]{$ 89.62 _{ - 0.36  } ^ {+ 0.25 }$ #1} 
\newcommand{\ab}[1][AU]{$ 0.179 _{ - 0.027  } ^ {+ 0.021 }$ #1}   
\newcommand{\densb}[1][$\mathrm{g\,cm^{-3}}$]{$ 2.11 _{ - 0.81 } ^ { + 0.74 }$  #1}
\newcommand{\mpb}[1][$M_\mathrm{J}$]{$0.387 _{ - 0.075  } ^ {+ 0.083 }$ #1} 
\newcommand{\rpb}[1][$R_\mathrm{J}$]{$0.808 _{ - 0.033  } ^ {+ 0.034 }$ #1}   
\newcommand{\denpb}[1][$\mathrm{g\,cm^{-3}}$]{$0.91 _{ - 0.20} ^ { + 0.24  }$ #1}
\newcommand{\Tequib}[1][K]{$ 565 _{ - 32  } ^ {+ 48 }$  #1}    
\newcommand{\ttotb}[1][hours]{$ 4.89 _{ - 0.22  } ^ {+ 0.08}$  #1}
\newcommand{\Tzeroc}[1][days]{$7332.4 _{ - 5.1 } ^ { + 5.5 } $#1} 
\newcommand{\Pc}[1][days]{$17.26 \pm { 0.12 } $ #1} 
\newcommand{\kc}[1][$m \, s^{-1}$]{$7.1 _{ - 5.0} ^ {+ 7.5 }$ #1} 
\newcommand{\Tzerod}[1][days]{$7321.3 \pm 2.2 $#1} 
\newcommand{\Pd}[1][days]{$8.63 \pm 0.06 $ #1} 
\newcommand{\kd}[1][$m \, s^{-1}$]{$10.6 _{ - 6.9  } ^ {+ 7.7 }$ #1} 
\newcommand{\qone}[1][]{ $0.37_{ - 0.13}^{ + 0.18 } $ #1}   
\newcommand{\qtwo}[1][]{ $0.48_{ - 0.16}^{ + 0.24 } $ #1}

\newcommand{\velFIES}[1][$\mathrm{km\,s^{-1}}$]{ $ -31.3575 \pm 0.0064 $ #1}
\newcommand{\velHARPSN}[1][$\mathrm{km\,s^{-1}}$]{ $ -31.1950_{ - 0.0128}^{ + 0.0122} $ #1}
\newcommand{\velHARPS}[1][$\mathrm{km\,s^{-1}}$]{ $ -31.1970 \pm 0.0093 $ #1}
\newcommand{\rvjitterFIES}[1][$\mathrm{m\,s^{-1}}$]{ $ 9.6_{ - 6.5}^{ + 9.8 } $ #1}    
\newcommand{\rvjitterHARPSN}[1][$\mathrm{m\,s^{-1}}$]{ $ 10.2_{ - 7.3}^{ + 15.8 } $ #1}    
\newcommand{\rvjitterHARPS}[1][$\mathrm{m\,s^{-1}}$]{ $ 15.4_{ - 7.6}^{ + 11.0 } $ #1}        

\newcommand{\cmassp}{$49^{+19}_{-17}\,M_{\oplus}$}

\newcommand\corot{\emph{\it CoRoT}}
\newcommand\kepler{\emph{{\it Kepler}}}
\newcommand\vsini{$v$\,sin\,$i_\star$}

\newcommand\vmic{$v_{\rm mic}$}
\newcommand\vmac{$v_{\rm mac}$}

\newcommand\teff{$T_{\rm eff}$}

\newcommand\logg{log\,{\it g$_\star$}}

\usepackage[T1]{fontenc}
\usepackage{ae,aecompl}

\usepackage{newtxtext,newtxmath}
\title[The transiting warm Jupiter K2-139\,b]{K2-139\,b: a low-mass warm Jupiter on a 29-day orbit transiting an active K0\,V star}

\author[O. Barrag\'an et al.]{
O.~Barrag\'{a}n$^{1}$\thanks{E-mail: oscar.barraganvil@edu.unito.it},
D.~Gandolfi$^{1}$,
A.~M.~S.~Smith$^{2}$,
H.~J.~Deeg$^{3,4}$,
M.~C.~V.~Fridlund$^{5,6}$,
\newauthor
C.~M.~Persson$^{5}$,
P.~Donati$^{7}$,
M.~Endl$^{8}$,
Sz.~Csizmadia$^2$,
S.~Grziwa$^{9}$,
D.~Nespral$^{3,4}$,
\newauthor
A.~P.~Hatzes$^{10}$,
W.~D.~Cochran$^{8}$,
L.~Fossati$^{11}$,
S. S.~Brems$^{12}$,
J.~Cabrera$^{2}$,
F.~Cusano$^{7}$,
\newauthor
Ph.~Eigm\"{u}ller$^{1}$,
C.~Eiroa$^{13}$,
A.~Erikson$^{2}$,
E.~Guenther$^{10}$,
J.~Korth$^{9}$,
D.~Lorenzo-Oliveira$^{14}$,
\newauthor
L.~Mancini$^{15,16,17}$,
M.~P\"{a}tzold$^{9}$,
J.~Prieto-Arranz$^{3,4}$,
H.~Rauer$^{2,18}$
I.~Rebollido$^{13}$,
\newauthor
J.~Saario$^{19}$ and
O.V.~Zakhozhay$^{15,20}$ 
\\
~\\
$^{1}$Dipartimento di Fisica, Universit\'{a} di Torino, via P. Giuria 1, 10125 Torino, Italy\\
$^{2}$Institute of Planetary Research, German Aerospace Center, Rutherfordstrasse 2, 12489 Berlin, Germany\\
$^{3}$Instituto de Astrof\'{i}sica de Canarias, 38205 La Laguna, Tenerife, Spain\\
$^{4}$Departamento de Astrof\'{i}sica, Universidad de La Laguna, 38206 La Laguna, Tenerife, Spain\\
$^{5}$Department of Earth and Space Sciences, Chalmers University of Technology, Onsala Space Observatory, 439 92 Onsala, Sweden\\
$^{6}$Leiden Observatory, University of Leiden, PO Box 9513, 2300 RA, Leiden, The Netherlands\\
$^{7}$INAF - Osservatorio Astronomico di Bologna, Via Ranzani, 1, 20127, Bologna, Italy\\
$^{8}$Department of Astronomy and McDonald Observatory, University of Texas at Austin, 2515 Speedway, Stop C1400, Austin, TX 78712, USA\\
$^{9}$Rheinisches Institut f\"{u}r Umweltforschung an der Universit\"{a}t zu K\"{o}ln, Aachener Strasse 209, 50931 K\"{o}ln, Germany\\
$^{10}$Th\"{u}ringer Landessternwarte Tautenburg, Sternwarte 5, 07778 Tautenburg, Germany\\
$^{11}$Space Research Institute, Austrian Academy of Sciences, Schmiedlstrasse 6, A-8041 Graz, Austria\\
$^{12}$Landessternwarte, Zentrum f\"{u}r Astronomie der Universit\"{a}t Heidelberg, K\"{o}nigstuhl 12, 69117 Heidelberg, Germany \\
$^{13}$Departamento F\'isica Te\'orica, Unversidad Aut\'onoma de Madrid, Cantoblanco, 28049 Madrid, Spain \\
$^{14}$Universidade de S\~ao Paulo, Departamento de Astronomia do IAG/USP, Rua do Mat\~ao 1226, Cidade Universit\'aria, 05508-900 S\~ao Paulo, SP, Brazil \\
$^{15}$Max-Planck-Institut f\"{u}r Astronomie,\ K\"{o}nigstuhl  17, D-69117 Heidelberg, Germany \\
$^{16}$Department of Physics, University of Rome Tor Vergata, Rome \\
$^{17}$INAF - Astrophysical Observatory of Turin, Turin \\
$^{18}$Center for Astronomy and Astrophysics, TU Berlin, Hardenbergstr. 36, 10623 Berlin, Germany\\
$^{19}$Nordic Optical Telescope, Apartado 474, E-38700 Santa Cruz de La Palma, Spain \\
$^{20}$Main Astronomical Observatory, National Academy of Sciences of the Ukraine, 27 Akademika Zabolotnoho St. 03143, Kyiv, Ukraine \\
}

\date{Last updated ; in original form }

\pubyear{2017}

\begin{document}
\label{firstpage}
\pagerange{\pageref{firstpage}--\pageref{lastpage}}
\maketitle

\begin{abstract}
We announce the discovery of K2-139\,b (EPIC\,218916923\,b), a transiting warm-Jupiter ($T_\mathrm{eq}$=547$\pm$25\,K) on a 29-day orbit around an active (log\,$R^\prime_\mathrm{HK}$\,=\,$-$4.46\,$\pm$\,0.06) K0\,V star in \emph{K2} Campaign~7. We~derive the system's parameters by combining the \emph{K2} photometry with ground-based follow-up observations. With a mass of~\mpb\ and radius of \rpb, K2-139\,b is one of the transiting warm Jupiters with the lowest mass known to date. The planetary mean density of \denpb can be explained with a core of $\sim$50\,$M_\oplus$. Given the brightness of the host star ($V$\,=\,11.653\,mag), the relatively short transit duration ($\sim$5~hours), and the expected amplitude of the Rossiter-McLaughlin effect ($\sim$25\,\ms), K2-139\ is an ideal target to measure the spin-orbit angle of a planetary system hosting a warm Jupiter.
\end{abstract}


%%%%%%%%%%%%%%%%%%%%%%%%%%%%%%%%%%%%%%%%%%%%%%%%%%

%%%%%%%%%%%%%%%%% BODY OF PAPER %%%%%%%%%%%%%%%%%%

% The MNRAS class isn't designed to include a table of contents, but for this document one is useful.
% I therefore have to do some kludging to make it work without masses of blank space.
%\begingroup
%\let\clearpage\relax
%\tableofcontents
%\endgroup
%\newpage

\section{Introduction}
\label{sec:introduction}

Gas-giant planets \citep[$M_\mathrm{p}$\,$\gtrsim$\,0.3\,$M_{\mathrm{Jup},}$][]{Hatzes2015} with orbital periods ranging between $\sim$10 and 100 days are called warm Jupiters. They mark the transition between hot Jupiters (giant planets with orbital period between $\sim$1 and 10 days) and Jupiter analogues (orbital period longer than 100 days). They seem to be less common than hot Jupiters and their formation scenario is still under debate \citep[e.g.,][]{2016MNRAS.455.1538F,Boley2016}. Whereas it is commonly accepted that hot Jupiters did not form \emph{in situ} \citep[e.g.,][]{2012ARA&A..50..211K}, but rather formed beyond the snow line and then migrated inwards to their current position, it has been recently proposed that warm Jupiters might have formed \emph{in situ} \citep[e.g.,][]{Boley2016,Huang2016}. 

Eighty warm Jupiters have been discovered so far from both ground- \citep[e.g.,][]{daSilva2007,Brahm2016,Jenkins2017} and space-based surveys \citep[e.g.,][]{Deeg2010, Saad-Olivera2017,2017MNRAS.464.2708S}. About thirty are known to transit their parent star and only thirteen have masses and radii known with a precision better than 25\,\%\footnote{Source: \url{http://exoplanet.eu}, as of January 2017.}. They have been detected both in low-eccentricity orbits  \citep[$e$\,$\lesssim$\,0.4, e.g.,][]{Brahm2016,Niedzielski2016,2017MNRAS.464.2708S}, as well as in highly eccentric orbits \citep[e.g.,][]{2012ApJ...761..163D,Ortiz2014}. 
\citet{Dong2014a} found that warm Jupiters with high eccentricities ($e$\,$\gtrsim$\,0.4) tend to have a massive planetary/stellar companion in a long period orbit. The architectures of these systems suggest that eccentric warm Jupiters might have reached their current positions via high-eccentricity migration excited by the outer companion \citep{Dong2014a}. On the other hand, warm Jupiters with no detected Jovian companion tend to have lower eccentricities peaked around 0.2. This suggests that two different types of warm Jupiters might exist: those formed via high-eccentricity migration and those formed \emph{in situ}. Alternatively, warm Jupiters in low-eccentricity orbits can also result from disc-driven migration from the outer region of the system \citep{2012ARA&A..50..211K}.

\cite{Petrovich2016} studied the possibility that warm Jupiters are undergoing secular eccentricity oscillations induced by an outer companion in an eccentric and/or mutually inclined orbit. Their model suggests that high-eccentricity migration can account for most of the hot Jupiters, as well as for most of the warm Jupiters with $e$\,$\gtrsim$\,0.4. However, it cannot account for the remaining population of low-eccentricity warm Jupiters, which must have undergone a different formation mechanism.
The low efficiency to generate warm Jupiters in nearly circular orbits via high-eccentricity migration has been corroborated by \citet{Hamers2016} and \citet{Antonini2016} using numerical simulations. 

In order to test different planet formation mechanisms, we need to characterize the population of warm Jupiters in terms of planetary mass, radius and orbital parameters. 
We herein present the discovery of \pname\ (\sname b), a transiting warm Jupiter ($M_\mathrm{p}=$ \mpb, $R_\mathrm{p}=$ \rpb) in a 29-day orbit around an active K0\,V star that has been photometrically monitored by the \emph{K2} space-mission during its Campaign 7. We combine the \emph{K2} photometry with ground-based imaging and high-precision radial velocity measurements to confirm the planet and derive the main parameters of the system.

%---------------------------------------------------------------------------

\section{\emph{K2} photometry}
\label{sec:k2}

\emph{K2} Campaign 7 was performed between 2015 October 04 UT and  2015 December 26 UT\footnote{See \url{http://keplerscience.arc.nasa.gov/k2-fields.html}.}. 
The {\it Kepler} spacecraft was pointed at coordinates $\alpha=19^{\mathrm{h}}11^{\mathrm{m}}19^{\mathrm{s}}$, $\delta=-23^{\circ}21^\prime36^{\prime\prime}$. \emph{K2} observed simultaneously 13\,469 sources in long cadence mode ($\sim$30 minute integration time) and 72 objects in short cadence mode ($\sim$1~minute integration time), leading to a total of  13\,541 light curves.

For the detection of transiting planet candidates, we used the \emph{K2} Campaign 7 light curves\footnote{Publicly available at \url{https://www.cfa.harvard.edu/~avanderb/allk2c7obs.html}.} extracted by \cite{Vanderburg2014}. We analyzed the light curves using the \texttt{DST} algorithm \citep{Cabrera2012} and the \texttt{EXOTRANS} pipeline \citep{Grziwa2012,Grziwa2016}.
Both codes have been used extensively on \corot\ \citep{Carpano2009,Cabrera2009,Erikson2012,Carone2012,Cavarroc2012} and \kepler\ \citep[][]{Cabrera2014,Grziwa2016} data. These search algorithms detect periodic patterns in time series photometric data. \texttt{DST} uses an optimized transit shape with the same number of free parameters as for the \texttt{BLS} algorithm \citep[Box-fitting Least Squares;][]{Kovacs2002}, and it also implements better statistics for signal detection. \texttt{EXOTRANS} uses the \texttt{BLS} algorithm combined with the wavelet-based filter technique \texttt{VARLET} \citep{Grziwa2016}, diminishing the effects of stellar variability and data discontinuities.

\begin{table}
\caption{Main identifiers, coordinates, optical and infrared magnitudes, and proper motion of \ssname.  \label{tab:parstellar} 
}
\begin{center}
\begin{tabular}{lcc} 
\hline
\hline
\noalign{\smallskip}
Parameter & Value &  Source \\
\noalign{\smallskip}
\hline
\noalign{\smallskip}
\multicolumn{3}{l}{\emph{Main Identifiers}} \\
\noalign{\smallskip}
TYC  & 6300-2008-1 & Tycho \\
EPIC & 218916923  & EPIC \\
UCAC & 361-185490 & EPIC \\
2MASS & 19161596-1754384 & EPIC \\
\noalign{\smallskip}
\hline
\noalign{\smallskip}
\multicolumn{3}{l}{\emph{Equatorial coordinates}} \\
\noalign{\smallskip}
$\alpha$(J2000.0) & $19^\mathrm{h}16^{\mathrm{m}}15.967^{\mathrm{s}}$ & 2MASS \\
$\delta$(J2000.0) & -17$^{\circ}$54$^\prime$38.48${\arcsec}$ & 2MASS \\
\noalign{\smallskip}
\hline
\noalign{\smallskip}
\multicolumn{3}{l}{\emph{Magnitudes}} \\
$B$ & 12.433$\pm$0.205 & EPIC \\
$V$ & 11.653$\pm$0.137 & EPIC \\
$g$ & 12.049$\pm$0.010 & EPIC \\
$r$ & 11.400$\pm$0.010 & EPIC \\
$J$ & 10.177$\pm$0.022 & 2MASS \\
$H$ &  9.768$\pm$0.022 & 2MASS \\
$K$ &  9.660$\pm$0.023 & 2MASS \\
$W1$ & 9.598$\pm$0.024 & WISE \\
$W2$ & 9.684$\pm$0.020 & WISE \\
$W3$ & 9.593$\pm$0.043 & WISE \\
$W4$ & 8.487           & WISE \\
\noalign{\smallskip}
\hline
\noalign{\smallskip}
\multicolumn{3}{l}{\emph{Proper motions}} \\
$\mu_{\alpha} \cos \delta$ (mas \ yr$^{-1}$) & $38.584 \pm 3.907 $ & Gaia\\
$\mu_{\delta} $ (mas \ yr$^{-1}$) & $-9.837 \pm 3.534$ & Gaia \\
\noalign{\smallskip}
\hline
\end{tabular}
\end{center}
\begin{tablenotes}\footnotesize
 \item \emph{Note} -- Values of fields marked with EPIC are taken from the Ecliptic Plane Input Catalog, available at \url{http://archive. stsci.edu/k2/epic/search.php}. Values marked with Gaia, 2MASS, and WISE are from \citet{Fabricius2016},  \citet{Cutri2003}, and \citet{2012wise.rept....1C}, respectively. The WISE W4 magnitude is an upper limit.
\end{tablenotes}
\end{table}

\begin{figure*}
\includegraphics[width=1.0\textwidth,trim={0 0.35cm 0 0},clip]{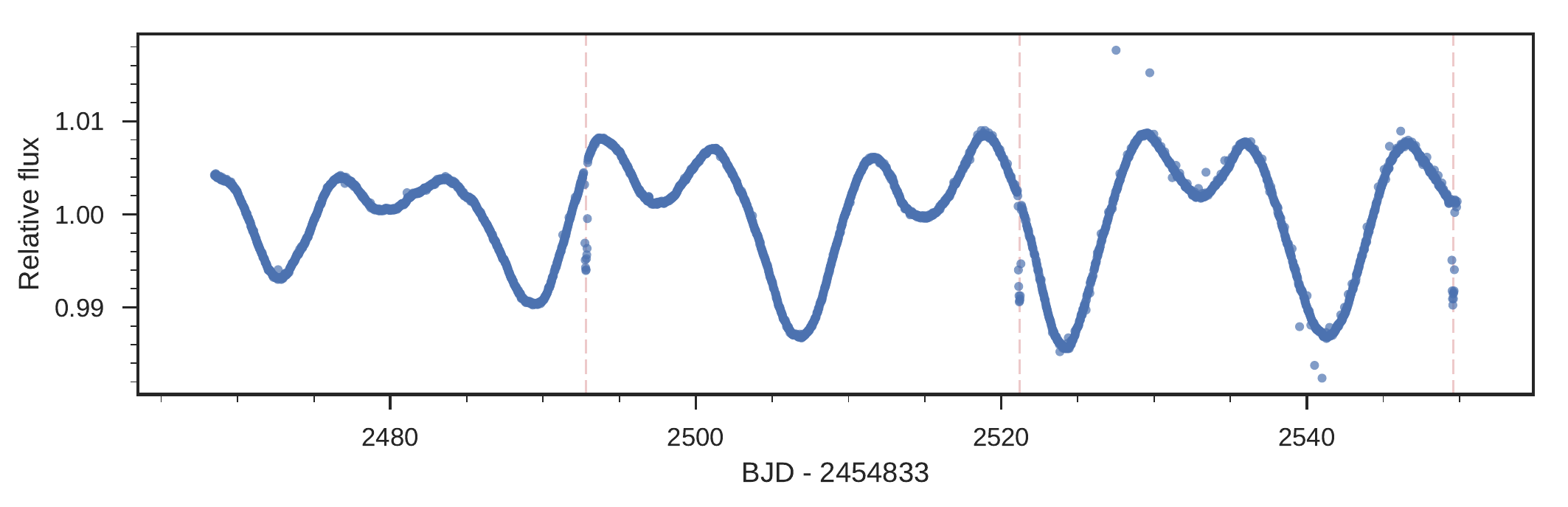}
\caption{\emph{K2} Light curve for \ssname\ as extracted by \citet{2016AJ....152..100L}. The positions of the 3 observed transits are marked with vertical dashed lines. 
\label{fig:lc}}
\end{figure*}

We detected a periodic transit-like signal associated with the star \sname\ with both \texttt{DST} and \texttt{EXOTRANS}. As a sanity check, we downloaded the EVEREST light curve of \sname\ \citep{2016AJ....152..100L} and detected the same signal. We note that \citet{Vanderburg2014} and \citet{2016AJ....152..100L} used the same mask to extract the time-series data from the raw \emph{K2} images. \sname\ was proposed for \emph{K2} observations by programs GO7086 (P.I. Thompson), GO7030 (P.I. Howard) and GO7087 (P.I. Dragomir). We will hereafter refer to the star and its transiting planet as \ssname\ and \pname, respectively.

We searched the \citet{Vanderburg2014}'s light curve for odd-even transit depth variation and secondary eclipse that might hint to a binary scenario making the system a likely false positive. None of them were significantly detected. The depth of the odd/even transits agrees within the 1-$\sigma$ uncertainty of $1.6 \times 10^{-3}$, whereas the 3-$\sigma$ upper limit of the occultation depth is $7.9\times 10^{-5}$, both respect to the normalized flux.
We proceeded to more detailed fitting of the light curve, as well as ground-based imaging (Sect.~\ref{sec:imaging}) and spectroscopic observations (Sect.~\ref{sec:rv}). The main identifiers, coordinates, optical and infrared magnitudes, and proper motions of the star are listed in Table~\ref{tab:parstellar}. We display the EVEREST \emph{K2} light curve of \ssname\ in Fig.~\ref{fig:lc}.

%---------------------------------------------------------------------------

\section{ALFOSC imaging}
\label{sec:imaging}

\emph{K2} Campaign 7 is projected close to the galactic center and thus in a relatively crowded stellar region. In order to estimate the contamination factor arising from sources whose light leaks into the photometric masks used by \citet{Vanderburg2014} and \citet{2016AJ....152..100L}, we observed  \ssname\ on 13 September 2016 (UT) with the ALFOSC camera mounted at the Nordic Optical Telescope (NOT) of Roque de los Muchachos Observatory (La Palma, Spain). The sky conditions were photometric with excellent seeing conditions ($\sim$0.6\arcsec). We used the Bessel R-filter and acquired 16 images of 6 sec, 2 images of 20 sec, and 1 image of 120 sec. The data were bias subtracted and flat-fielded using dusk sky flats. Aperture photometry was then performed on all stars within the mask used in the extraction of the light curve by \citet{Vanderburg2014} and \citet{2016AJ....152..100L}.

\begin{figure}
\begin{center}
\includegraphics[width=0.35\textwidth]{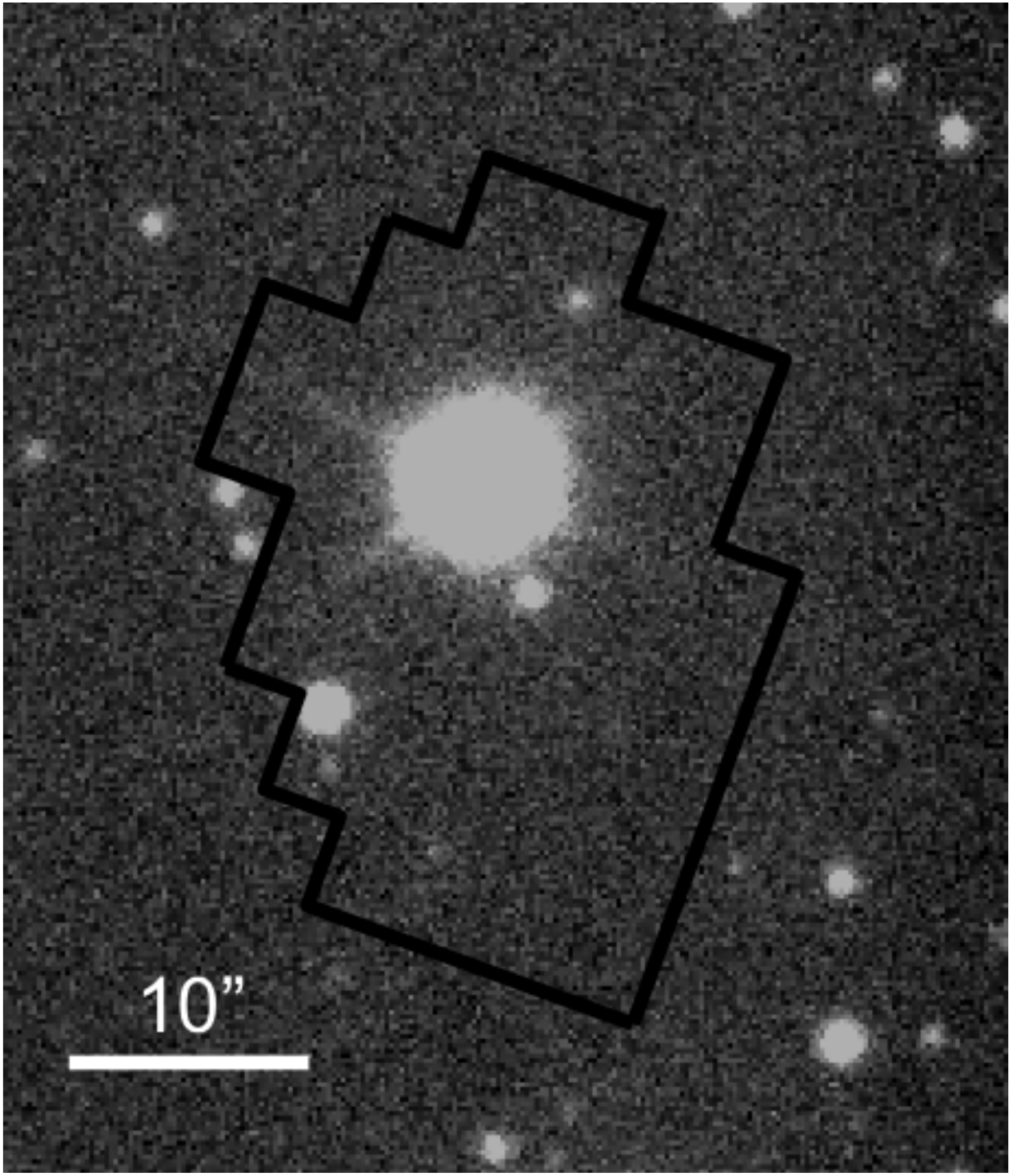}
\caption{ALFOSC Bessel R-band image of the sky region around \ssname. North is up and East is to the left. The target star is the brightest source in the middle. The solid black polygon marks the EVEREST photometric mask \citep{2016AJ....152..100L}. \label{fig:alfosc}
}
\end{center}
\end{figure}

Several fainter stars can be identified inside the photometric mask (Fig.~\ref{fig:alfosc}), of which the two brightest sources are also in the EPIC catalog with Kepler band magnitudes of $16.8$ and $18.4$. The closest detected source is a 6.8-mag fainter star at 3.8$^{\prime\prime}$ South of  \ssname. We can exclude stars as faint as $\sim$20\,mag at an angular distance larger than $\sim$0.6$\arcsec$ from \ssname. It is worth noting that the faintest star whose flux could account for the $\sim$1\% deep transit of  \ssname\ cannot be more than $\sim$5 mag fainter than our target. The summed flux of these faint stars amounts to 1.4$\pm$0.3\,\% of the total off-transit flux within the aperture.
We subtracted this contamination flux from the EVEREST \emph{K2} light curve prior to performing the joint analysis presented in Sect.~\ref{sec:joint}.
 
%---------------------------------------------------------------------------

\section{High-resolution spectroscopy}
\label{sec:rv}

\begin{table*}
\centering
%\begin{center}
\caption{Radial velocity measurements and activity indexes of \ssname. \label{rvs}}
\begin{tabular}{cccrccc}
\hline
\hline
BJD$_\mathrm{TDB}$ & RV & $\sigma_{\mathrm{RV}}$ & CCF BIS & CCF FWHM & log\,$R^\prime_\mathrm{HK}$ & $\sigma_{log\,R^\prime_\mathrm{HK}}$\\
$-{2\,450\,000}$ & (km s$^{-1}$) & (km s$^{-1}$)  & (km s$^{-1}$) & (km s$^{-1}$) \\
\hline
\noalign{\smallskip}
\multicolumn{2}{l}{FIES} \\
 7565.656116 & $-$31.3755 & 0.0160 &    0.0119 & 12.1638 & \dotfill & \dotfill  \\
 7568.556388 & $-$31.3503 & 0.0155 &    0.0129 & 12.1080 & \dotfill & \dotfill  \\
 7569.567239 & $-$31.3317 & 0.0153 &    0.0264 & 12.1590 & \dotfill & \dotfill  \\
 7570.606019 & $-$31.3473 & 0.0136 &    0.0098 & 12.1547 & \dotfill & \dotfill  \\
 7572.576513 & $-$31.3357 & 0.0133 &    0.0107 & 12.1226 & \dotfill & \dotfill  \\
 7574.529831 & $-$31.3466 & 0.0101 &    0.0072 & 12.1158 & \dotfill & \dotfill  \\
 7576.536114 & $-$31.2990 & 0.0136 &    0.0016 & 12.1254 & \dotfill & \dotfill  \\
 7579.547224 & $-$31.3441 & 0.0139 & $-$0.0015 & 12.1284 & \dotfill & \dotfill  \\
 7585.551244 & $-$31.3706 & 0.0111 &    0.0084 & 12.1410 & \dotfill & \dotfill  \\
 7589.540362 & $-$31.3913 & 0.0143 &    0.0130 & 12.1236 & \dotfill & \dotfill  \\
\hline
\noalign{\smallskip}
\multicolumn{2}{l}{HARPS} \\
7569.714094 & $-$31.1633 & 0.0032 &  0.0144 & 7.4922 & -4.552 & 0.028 \\
7587.830287 & $-$31.2146 & 0.0052 &  0.0142 & 7.4843 & -4.578 & 0.060 \\
7589.523734 & $-$31.2116 & 0.0049 &  0.0131 & 7.5051 & -4.596 & 0.042 \\
7610.717929 & $-$31.2217 & 0.0028 &  0.0045 & 7.4363 & -4.588 & 0.025 \\
7619.531746 & $-$31.2190 & 0.0031 & $-$0.0146 & 7.4440 & -4.498 & 0.021 \\
7620.682635 & $-$31.2049 & 0.0069 &  0.0069 &  7.4263 & -4.455 & 0.052 \\
\hline
\noalign{\smallskip}
\multicolumn{2}{l}{HARPS-N} \\
7586.621783 & $-$31.2048 & 0.0029 & 0.0103 & 7.4501 & -4.461 & 0.018 \\
7587.603577 & $-$31.2141 & 0.0038 & 0.0072 &  7.4396 & -4.476 & 0.025 \\
7605.429766 & $-$31.1683 & 0.0050 & $-$0.0003 &  7.4336 & -4.479 & 0.040 \\
\hline
\end{tabular}
%\end{center}
\end{table*}

In June and August 2016 we obtained two reconnaissance spectra of \ssname\ with the Tull spectrograph \citep{Tull1995} at the 2.7-m telescope at McDonald Observatory (Texas, USA). The high resolution ($R \approx 60\,000$) spectra have a signal-to-noise ratio of $\sim$30 per pixel at 5500\AA. We reduced the data using standard \texttt{IRAF} routines and derived preliminary spectroscopic parameters using our code \texttt{Kea} \citep{Endl2016}. The results from both spectra are nearly identical and reveal a star with \teff\,=\,5500\,$\pm$\,100~K, \logg\,=\,4.65\,$\pm$\,0.12~(cgs), [Fe/H]=+0.11\,$\pm$\,0.12~dex, and a slow projected rotational velocity of \vsini\,$\approx$\,2~\kms.

The high-precision radial velocity follow-up of \ssname\ was started in June 2016 with the FIbre-fed \'Echelle Spectrograph \citep[FIES;][]{Frandsen1999,Telting2014} mounted at the 2.56-m Nordic Optical Telescope (NOT). The observations were carried out as part of the OPTICON and CAT observing programs 16A/055, P53-201, and P53-203. We used the \emph{high-res} mode, which provides a resolving power of $R$\,$\approx$\,$67\,000$ in the whole visible spectral range ($3700-7300$\,\AA). The exposure time was set to 2100\,--\,3600 sec, based on sky conditions and observing scheduling constraints. Following the observing strategy outlined in \citet{Buchhave2010} and \citet{Gandolfi2015}, we traced the RV drift of the instrument by acquiring long-exposed (T$_\mathrm{exp}$\,$\approx$\,35 sec) ThAr spectra immediately before and after the target observations. The typical RV drift measured between two ThAr spectra bracketing a 2100\,--\,3600 sec science exposure is about 50\,--\,80~\ms. A linear interpolation of the RV drift to the mid-time of the science exposure allows us to achieve a radial velocity zero-point stability of about 5\,--\,6~\ms, which is 2\,--\,3 times smaller than the nominal error bars listed in Table~\ref{rvs}. The data reduction uses standard \texttt{IRAF} and \texttt{IDL} routines. The signal-to-noise (S/N) ratio of the extracted spectra is $\sim$30\,--\,40 per pixel at 5500~\AA. Radial velocity measurements were extracted via multi-order cross-correlation with the RV standard star HD\,182572, observed with the same instrument set-up as \ssname.

We also observed \ssname\ in July, August, and September 2016 with the HARPS \citep{Mayor2003} and HARPS-N \citep{Cosentino2012} spectrographs mounted at the ESO 3.6-m Telescope of La Silla Observatory (Chile) and at the 3.58-m Telescopio Nazionale Galileo (TNG), of Roque de los Muchachos observatory (La Palma, Spain), respectively. Both instruments provide a resolving power of $R$\,$\approx$\,$115\,000$ in the wavelength range $\sim$3800\,--\,6900~\AA. The observations were performed as part of the ESO and TNG observing programs 097.C-0948 and A33TAC\_15, respectively. The exposure time was set to 1800 sec, leading to a S/N ratio of $\sim$35 on the extracted spectra. We reduced the data using the dedicated HARPS and HARPS-N pipelines and extracted the RVs by cross-correlation with a G2 numerical mask.

The FIES, HARPS, and HARPS-N RVs are listed in Table~\ref{rvs} along with the bisector span (BIS) and the full width at half maximum (FWHM) of the cross-correlation function (CCF). Time stamps are given in barycentric Julian date in barycentric dynamical time (BJD$_\mathrm{TDB}$). For the HARPS and HARPS-N data we also provide the Ca\,{\sc ii} H\,\&\,K chromospheric activity index log\,$R^\prime_\mathrm{HK}$. We did not measure log\,$R^\prime_\mathrm{HK}$ from the FIES spectra because of the poor S/N ratio at wavelengths shorter than 4000\,\AA.

%---------------------------------------------------------------------------

\section{Stellar parameters}
\label{sec:spec}

\subsection{Spectral analysis}

We derived the spectroscopic parameters of \ssname\ from the co-added FIES spectra. The stacked FIES data have a S/N ratio of $\sim$110 per pixel at 5500~\AA. We adopted three different methods. For each method, results are reported in Table~\ref{tab:spec_param}.

\begin{table*}%[!th]
\begin{center}
\caption{Spectroscopic parameters of \ssname\ as derived using the three methods described in Sect~\ref{sec:spec}.\label{tab:spec_param}}
\begin{tabular}{lcccccc}
\hline
\hline
\noalign{\smallskip}
Method &  \teff\ (K) & \logg\ (cgs) &  [Fe/H] (dex) &  \vmic\ (\kms) & \vmac\ (\kms) &  \vsini\ (\kms) \\
\hline
\noalign{\smallskip}
\multicolumn{3}{l}{\emph{Adopted spectroscopic parameters}} \\
 \noalign{\smallskip}
Method 1 & 5340$\pm$110 & 4.50$\pm$0.09 & 0.22$\pm$0.08 & 0.9$\pm$0.1 & 2.5$\pm$0.6 & 2.8$\pm$0.6\\
\noalign{\smallskip}
\hline
\noalign{\smallskip}
Method 2 & 5185$\pm$100 & 4.53$\pm$0.10 & 0.20$\pm$0.10 & 0.8$\pm$0.1 & 2.4$\pm$0.5 & 3.0$\pm$0.5\\
Method 3 & 5343$\pm$99  & 4.58$\pm$0.21 & 0.21$\pm$0.10 & 0.9$\pm$0.1 & -- & -- \\
\hline
\end{tabular}
\end{center}
\end{table*}

\emph{First method}. The technique fits spectral features that are sensitive to different photospheric parameters. It uses the stellar spectral synthesis program \texttt{Spectrum} \citep{Gray1999} to compute synthetic spectra from \texttt{ATLAS\,9} model atmospheres \citep{Castelli2004}. Microturbulent (\vmic) and macroturbulent (\vmac) velocities are derived from the calibration equations of \citet{Bruntt2010} and \citet{Doyle2014}. We used the wings of the H$_\alpha$ and H$_\beta$ lines to estimate the effective temperature (\teff), and the Mg\,{\sc i}~5167, 5173, and 5184~\AA, Ca\,{\sc i}~6162 and 6439~\AA, and the Na\,{\sc i}~D lines to determine the surface gravity \logg. We simultaneously fitted different spectral regions to measure the iron abundance [Fe/H]. The projected rotational velocity \vsini\ was determined by fitting the profile of many isolated and unblended metal lines.

\emph{Second method}. It relies on the use of the spectral analysis package Spectroscopy Made Easy  \citep[\texttt{SME};][]{vp96,vf05}. For a set of given stellar parameters, \texttt{SME} calculates synthetic spectra and fits them to high-resolution observed spectra using a chi-squared minimization procedure. We used \texttt{SME} version 4.4.3 and \texttt{ATLAS\,12} model spectra \citep{Kurucz2013}. We adopted the same calibration equation as described in the first method to determine \vmic\ and \vmac. Effective temperature is derived from the H$_\alpha$ wings; \logg\ from the Ca\,{\sc i}~6102, 6122, 6162, and 6439~\AA\ lines; [Fe/H] and \vsini\ from isolated iron lines.

\emph{Third method}. It uses the classical equivalent width (EW) method adopting the following criteria: {\it i}) \teff\ is obtained by removing trends between abundance of the chemical elements and the respective excitation potentials; {\it ii}) \logg\ is optimised by assuming the ionisation equilibrium condition, i.e., by requiring that for a given species, the same abundance (within the uncertainties) is obtained from lines of two ionisation states (typically, neutral and singly ionised lines); {\it iii}) \vmic\ is set by minimising the slope of the relationship between abundance and the logarithm of the reduced EWs. The equivalent widths of Fe\,{\sc i} and Fe\,{\sc ii} lines are measured using the code \texttt{DOOp} \citep{Cantat-Gaudin2014}, a wrapper of \texttt{DAOSPEC} \citep{Stetson2008}. The stellar atmosphere parameters are derived with the program \texttt{FAMA} \citep{Magrini2013}, a wrapper of \texttt{MOOG} \citep{Sneden2012}. We used the public version of the atomic data prepared for the {\it Gaia}-ESO Survey \citep{Heiter2015} and based on the \texttt{VALD3} data \citep{Ryabchikova2011}. We used $\sim$200 Fe\,{\sc i} lines and $\sim$10 Fe\,{\sc ii} lines for the determination of the stellar parameters.

The three methods provide consistent results within the 1-$\sigma$ error bars (Table~\ref{tab:parameters}). While we have no reason to prefer one technique over the other, we adopted the parameter estimates of the first method, i.e., \teff\,=\,$5340\pm110$~K, \logg\,=\,$4.50\pm0.09$\,(cgs), [Fe/H]\,=\,$0.22\pm0.08$\,dex, \vmic\,=\,$0.9\pm0.1$\,\kms, \vmac\,=\,$2.5\pm0.6$\,\kms\ and \vsini\,=\,$2.8\pm0.6$\,\kms. As a sanity check, we also analyzed the HARPS and HARPS-N data and obtained consistent results but with larger error bars, owing to the lower S/N ratio of the co-added HARPS and HARPS-N spectra compared to that of the co-added FIES data. Using the \cite{Boyajian2013}'s calibration (see their Table 6), the effective temperature of \ssname\ defines the spectral type of the host star as K0\,V.

\subsection{Interstellar extinction}
\label{sec:Av}

We measured the visual reddening ($A_\mathrm{V}$) of \ssname\ following the technique described in \citet{Gandolfi2008}. We fitted the spectral energy distribution of the star to synthetic colors extracted from the \texttt{BT-NEXTGEN} model spectrum \citep{Allard2011} with the same photospheric parameters as the star. We adopted the extinction law of \citet{Cardelli1989} and assumed a normal value for the total-to-selective extinction, i.e., $R_\mathrm{V}$\,=\,$A_\mathrm{V}$/$E(B-V)$\,=\,3.1. We measured a visual extinction of $A_\mathrm{V}$\,=\,0.07\,$\pm$\,0.05~mag. This value is below the upper limit of  $A_\mathrm{V}$\,$\lesssim$\,0.3 mag extracted from the \citet{Schlegel1998}'s all-sky extinction map, corroborating our result.

\subsection{Rotational period}
\label{Sect:Rot}

The \emph{K2} light curve of \ssname\ displays periodic and quasi-periodic variations with a peak-to-peak photometric amplitude of $\sim$2\,\% (Fig.~\ref{fig:lc}). The late-type spectral type of the star suggests that the observed variability is due to Sun-like spots appearing and disappearing from the visible stellar disc as the star rotates around its axis. This is corroborated by the fact that \ssname\ is a chromospherically active star. The HARPS and HARPS-N spectra show clear emission components in the cores of the Ca\,{\sc ii} H\&K lines, from which we measured an average activity index of log\,$R^\prime_\mathrm{HK}$\,=\,$-4.46\pm0.06$\footnote{This value is corrected for the interstellar medium absorption, following the procedure described in \citet{Fossati2017} and using the measured stellar parameters and reddening. The correction is +0.06. The star is therefore slightly more active than what measured from the spectra.}.

The out-of-transit photometric variability observed in the light curve of \ssname\ is mainly due to two active regions located at opposite stellar longitudes, whose lifetime is longer than the duration of the \emph{K2} observations. Using the spots as tracers of stellar rotation and following the auto correlation function (ACF) technique described in \citet{2014ApJS..211...24M}, we estimated that the rotational period of the star is $P_\mathrm{rot}$\,=\,$17.24\pm0.12$~days. The Lomb-Scargle periodogram of the light curve shows its strongest peak at the same period confirming our results.

It is worth noting that the rotation period ($P_\mathrm{rot}$\,=\,$17.24\pm0.12$~days) and radius ($R_{\star}$=\sradius; see next section) of the host star translate into a maximum value for the projected rotational velocity of \vsini$_\mathrm{,max}$\,=\,2.53\,$\pm$\,0.10~\kms, which agrees with the spectroscopically derived \vsini\,=\,$2.8\,\pm\,0.6$~\kms, suggesting that the star is seen nearly equator-on ($i_\star \approx 90$\degr) and that the system might be aligned along the line-of-sight. 

\subsection{Stellar mass, radius and age}

We derived the stellar mass, radius, and age using the online interface for Bayesian estimation of stellar parameters available at the following web page: \url{http://stev.oapd.inaf.it/cgi-bin/param}. Briefly, the web tool interpolates onto \texttt{PARSEC} model isochrones \citep{Bressan2012}, the V-band apparent magnitude, effective temperature, metal content, and parallax. We used the V-band magnitude reported in Table~\ref{tab:parstellar} -- after correcting for interstellar reddening (Sect~\ref{sec:Av}) -- along with the effective temperature and metal content we derived in  Sect.~\ref{sec:spec}. The parallax was retrieved from the {\it Gaia}'s first data release \citep[px\,=\,$6.56\pm0.43$\,mas, $d$\,=\,152\,$\pm$\,10~pc][]{Fabricius2016}. We adopted the log-normal initial mass function from \citet{Chabrier2001}.

\ssname\ has a mass of $M_{\star}$=\smass\ and radius of $R_{\star}$=\sradius, corresponding to a surface gravity of \logg\,=\,4.503$\pm$0.035 (cgs), in excellent agreement with the spectroscopically derived value of \logg\,=\,4.50$\pm$0.09~(cgs; see Sect.~\ref{sec:spec}). The derived mean density $\rho_\star = 2.02\pm0.24$~g\,cm$^{-3}$ of \ssname\ is also consistent within 1-$\sigma$ with the density estimated by the modeling of the transit light curve ($\rho_\star$\,=\,\densb; see Sect.~\ref{sec:joint}).

The isochrones provide an age of 3.6$\pm$3.4~Gyr for \ssname. Using the equations given in \citet{Barnes2010a} and \citet{Barnes2010b}, the rotation period of 17.3~days (Sect~\ref{Sect:Rot}) implies a gyrochronological age of 1.8\,$\pm$\,0.3~Gyr.

%---------------------------------------------------------------------------

\section{Joint RV-transit fit}
\label{sec:joint}

We performed the joint fit to the photometric and RV data using the code \texttt{pyaneti}\footnote{Available at \url{https://github.com/oscaribv/pyaneti}.} \citep{pyaneti}, a Python/Fortran software suite based on Markov Chain Monte Carlo (MCMC) methods. 

The photometric data included in the joint analysis are subsets of the whole EVEREST \emph{K2} light curve. We used the EVEREST light curve because it provides a slightly better rms over the \citet{Vanderburg2014}'s data. We selected $\sim$10 hours of data-points around each of the 3 transits, which have a duration of $\sim$5 hours. We de-trended each individual transits with the code \texttt{exotrending}\footnote{Available at \url{https://github.com/oscaribv/exotrending}.} \citep{Barragan2017}, using a second-order polynomial fitted to the out-of-transit points. The fitted data include 12 points immediately before and after each transit, with the exception of the last transit for which only 9 data points are available. We removed the data points that are affected by stellar spot crossing events (see Sect.~\ref{sec:stellaractivity} for more details).

We fitted the RV data using a Keplerian model for the planet, along with two sine-like curves to account for the activity-induced RV (see next section for details). We adopted the limb-darkened quadratic law of \citet{2002ApJ...580L.171M} for the transit model. We adopted the Gaussian likelihood

\begin{equation}
    \mathcal{L} = 
%--------------------------
     \left[
    \prod_{i=1}^{n}
    \left\lbrace 2 \pi \left( \sigma_i^2 + \sigma_{\rm j}^2 \right)
    \right\rbrace^{- 1/2}
    \right]
    \exp
    \left\lbrace
    - \sum_{i=1}^{n} \frac{1}{2} \frac{ 
    \left( D_i - M_i \right)^2}{ \sigma_i^2 + \sigma_{\rm j}^2 }
    \right\rbrace, 
  \label{eq:eq1}
\end{equation}

where $n$ is the number of data points, $\sigma_i$ is the error associated to each data point $D_i$, $M_i$ is the model associated to a given $D_i$ and $\sigma_{\rm j}$ is an extra noise term, sometime referred as jitter.

The sampling method and fitted parameters are the same as in \cite{Barragan2016}. Details on the adopted priors are given in Table~\ref{tab:parameters}.
Following \citet{Kipping2010}, we super-sampled the light curve model using 10 subsamples per \emph{K2} exposure to account for the long-cadence acquisition. The parameter space was explored with 500 independent chains created randomly inside the prior ranges. The chain convergence was analyzed using the Gelman-Rubin statistics. The burn-in phase uses $25,000$ more iterations with a thin factor of $50$. The posterior distribution of each parameter has $250,000$ independent data points.

%---------------------------------------------------------------------------

\section{Results and discussion}
\label{sec:res}

\subsection{Stellar activity modeling}
\label{sec:stellaractivity}

A simple Keplerian model provides a poor fit to the RV measurements with $\chi^2/{\rm dof} = 6.1$ (Table~\ref{tab:models}), suggesting that additional signals might be present in our Doppler data. Activity-induced RV variation is expected given the 2\,\% peak-to-peak photometric variability observed in the \emph{K2} light curve of \ssname\ (Fig.~\ref{fig:lc}) and the Ca\,{\sc ii} H\,\&\,K activity index of log\,$R^\prime_\mathrm{HK}$\,=\,$-4.46\pm0.06$ (Sect.~\ref{Sect:Rot}). The  \emph{K2} photometric variation corresponds to a spot filling factor of approximately 2\,\%, if this variation is due to cool starspots. We can use the empirical relationship relating spot coverage to RV amplitude from \citet{Saar1997} or \citet{Hatzes2002} to estimate the RV amplitude expected from spots. Using the projected rotational velocity of 2.8~\kms\ results in an RV semi-amplitude of $\approx$20--30\,\ms. The code SOAP\,2, designed to estimate the effect of active regions on photometric and spectroscopic measurements \citep{2014ApJ...796..132D}, provides consistent results.

In order to look for additional signals in our Doppler data, we performed a frequency analysis of the RV measurements and activity indicators. On one occasion\footnote{Epoch BJD=2457589.} \ssname\ was observed with FIES and HARPS-S nearly simultaneously (within less than 25 minutes). We used the two sets of measurements to estimate the RV, FWHM, and BIS offsets between the two instruments. We assumed no offset between HARPS-N and HARPS. While we acknowledge that this assumption is arbitrary, we note that the modeling of the RV data gives an offset of $\Delta$\,RV$_\mathrm{(HN-H)}$\,=\,0.002\,$\pm$\,0.0158\,\kms\ (Table~\ref{tab:parameters}), which is consistent with zero.

\begin{figure}
\begin{center}
\includegraphics[width=\columnwidth]{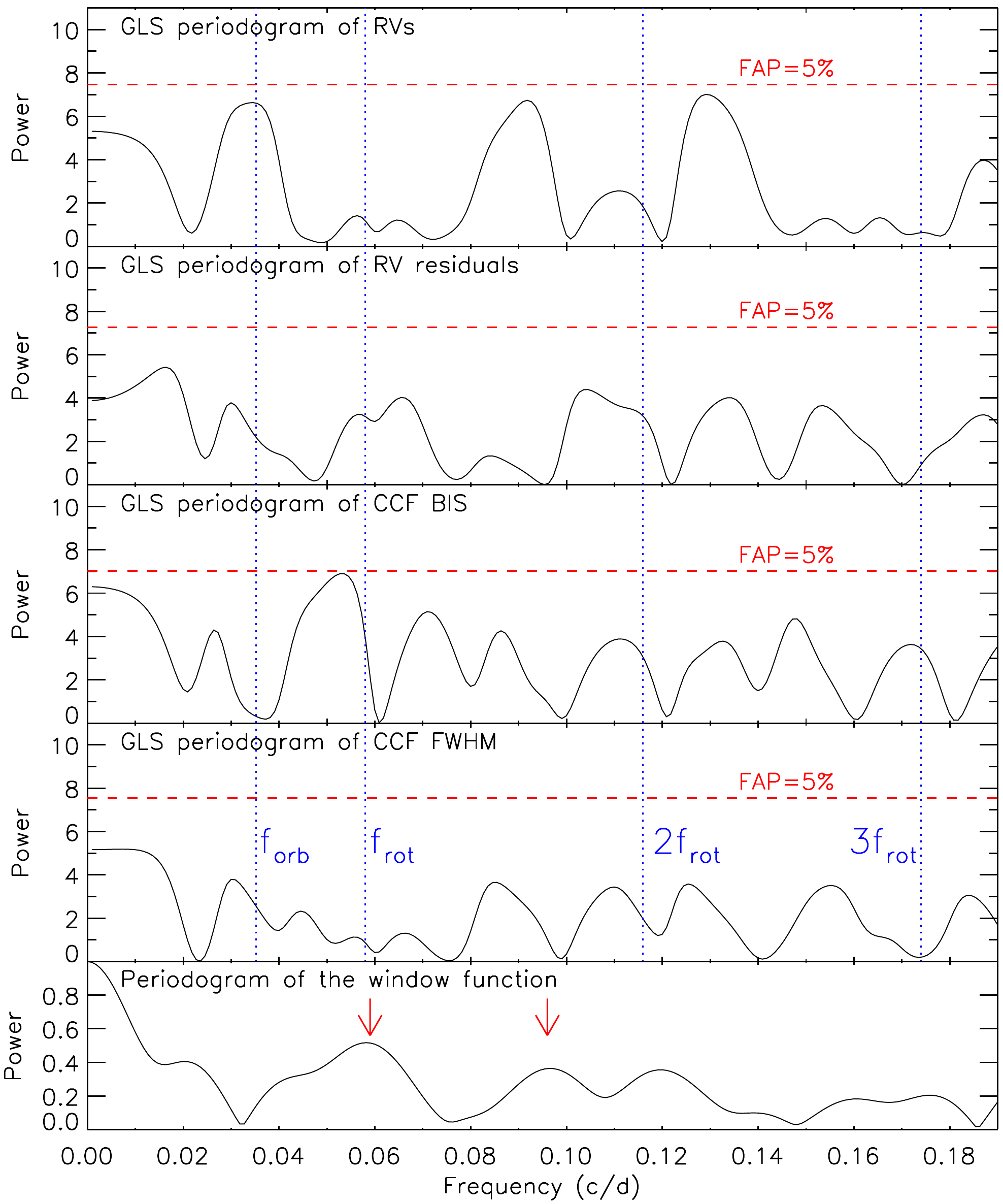}
\caption{Generalized Lomb-Scargle periodogram of the combined FIES, HARPS, and HARPS-N Doppler datasets. From top to bottom: the RV data, the RV residuals after subtracting the transiting planet signal, the BIS and FWHM of the CCF, and the window function. The dotted vertical blue lines mark the frequencies at the orbital period, as well as at the stellar rotation period and its first two harmonics. The dashed vertical red lines mark the 5\,\% false alarm probabilities as derived using the bootstrap method. The red arrows in the lower panel mark the two peaks presented in the main text.
\label{fig:gls}
}
\end{center}
\end{figure}

Figure~\ref{fig:gls} displays the generalized Lomb-Scargle periodograms \citep{2009A&A...496..577Z} of the combined datasets. From top to bottom the RV data (first panel), the RV residuals after subtracting the transiting planet signal (second panel), and the BIS (third panel) and FWHM (fourth panel) of the cross correlation function. The periodogram of the window function is shown in the lower panel. The dotted vertical blue lines mark the frequency at the orbital period of the planet (0.035~c/d), as well as the frequencies at the rotation period of the star (0.058~c/d) and its first two harmonics (0.116 and 0.174~c/d). 

The periodogram of the RV data (upper panel) shows a peak at the orbital frequency of the planet along with two additional peaks at 0.095 and 0.130~c/d. Since the periodogram of the window function shows two peaks at $\sim$0.060 and $\sim$0.095 c/d (lower panel, red arrows), we interpreted the 0.095 and 0.130 c/d peaks as the aliases of the orbital frequency\footnote{0.095\,=\,0.035\,+\,0.060 c/d and 0.130\,=\,0.035\,+\,0.095 c/d.}. We note also that periodogram of the BIS of the CCF displays peaks whose frequencies are close to the stellar rotation frequency and its first two harmonics. However, none of the peaks visible in the GLS periodograms of Fig.~\ref{fig:gls} has a false alarm probability (FAP)\footnote{We determined the FAP following the Monte Carlo bootstrap method described in \citet{Kuerster1997}.} lower than 5\,\%. Although our spectroscopic data show neither additional signals, we note that the semi-amplitude variation of the BIS and FWHM is expected to be $\lesssim$\,10-15\,\ms\ \citep{2014ApJ...796..132D}, which is comparable with the uncertainties of most of our measurements (Table~\ref{rvs}). The lack of significant peaks in the periodogram of the RV data and RV residuals, as well as in the periodogram of the activity indicators, could be explained by the limited number of available measurements and their uncertainties. We conclude that we cannot exclude the existence of spot-induced signals in our RV measurements.

Photometric and radial velocity variations due to rotational modulation can be complex with not only the rotational period $P_{\rm rot}$ present, but also its harmonics, e.g., $P_{\rm rot}$/2, $P_{\rm rot}$/3. Assuming that the surface structures responsible for this modulation (e.g., cool spots) are not evolving rapidly, then the simplest representation of the rotational modulation is through the Fourier components defined by the rotation period and its harmonics. Figure~\ref{fig:lc} shows that the evolution time-scale of the active regions in the stellar surface is longer than the 80-day duration of the {\it K2} campaign. Since our RV follow-up spans 55 days, we can assume that any activity-induced RV signal is coherent within our observing window. This approach has been used previously for other planetary systems orbiting active stars \citep[e.g.,][]{Pepe2013}.

The Fourier analysis of the {\it K2} light curve is the best way to measure the contribution of the rotation period and its harmonics to the quasi-periodic photometric variability of the star. We therefore analyzed the {\it K2} light curve using a pre-whitening procedure. That is, the dominant period was found, a sine-fit made to the data and subtracted, and additional periods searched in the residual data. We used the program \texttt{Period04} \citep{Lenz2005} for this procedure.

The dominant periods are $\sim$17.2 days, i.e., the rotation period of the star (Sect.~\ref{Sect:Rot}), and roughly the first four harmonics (i.e., 8.6, 5.7, 4.3, and 3.4 days). The 17.2- and 8.6-day periods have about the same amplitude, while the 5.7-day period ($P_{\rm rot}$/3) has 10\% of the main amplitude. The $P_{\rm rot}$/4 signal has only about 4\% of the main amplitude. The light curve analysis indicates that the signal due to rotational modulation can largely be represented by the rotational period ($P_{\rm rot}$) and its first harmonic ($P_{\rm rot}$/2) .

In order to test if the addition of RV sinusoidal signals at the stellar rotation period and its harmonics can account for the additional variation seen in our RV measurements, we compared different models by adding signals one by one. The first model (P0) includes only the planet signal, i.e., a Keplerian model fitted to the RV data using the same priors given in Table~\ref{tab:parameters}, but fixing epoch and period to the values derived by the transit modeling. The next model (P1) is obtained from P0 by adding a sinusoidal signal at the rotation period of the star ($P_{\rm rot}$). Models P2 includes the first harmonic of the rotation period ($P_{\rm rot}$/2), whereas model P3 account for the first ($P_{\rm rot}$/2) and second ($P_{\rm rot}$/3) harmonics. While adding sinusoidal signals, we fitted for their amplitudes, phases and periods.  We used flat priors for the phases and amplitudes (details in Table~\ref{tab:parameters}). We used a Gaussian prior for $P_{\rm rot}$ using the value and its uncertainty derived in Sect.~\ref{Sect:Rot}. The periods of the harmonic signals were left free to vary depending on the value assumed by $P_{\rm rot}$ at each step of the MCMC chains. In order to check if the RV variation induced by the planet is significant in our data set, we also performed the fit using models where the planetary signal was not included (models NP1 and NP2; see Table \ref{tab:models}).

Table~\ref{tab:models} shows the goodness of the fit for each model. The preferred model is P2 (planet plus 2 sinusoidal signals at $P_{\rm rot}$ and $P_{\rm rot}$/2) with the lowest Akaike Information Criteria (AIC) and maximum likelihood. This result is consistent with the Fourier analysis of the \emph{K2} light curve, which suggests that the major contribution to the photometric variations arises from the stellar rotation period and its first harmonic. Our analysis provides also additional evidence that the Doppler motion induced by the planet is present in our RV data set. First, the planet signal does not significantly vary for the P0, P1, P2 and P3 models (Table~\ref{tab:models}). Second, the models with no planetary signal (NP1 and NP2) provide a poor fit to the RV measurements (Table~\ref{tab:models}). 

\begin{figure*}
\includegraphics[width=0.95\textwidth]{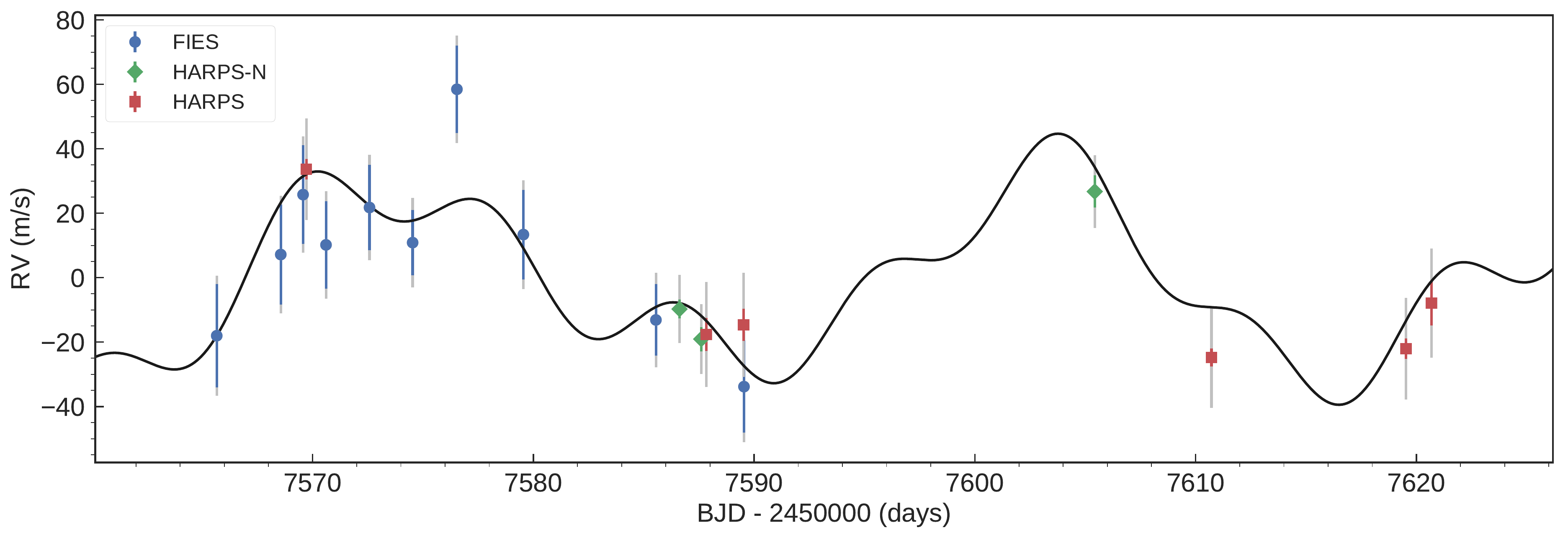} \\
\includegraphics[height=0.28\textwidth,trim={0 0.35cm 0 0},clip]{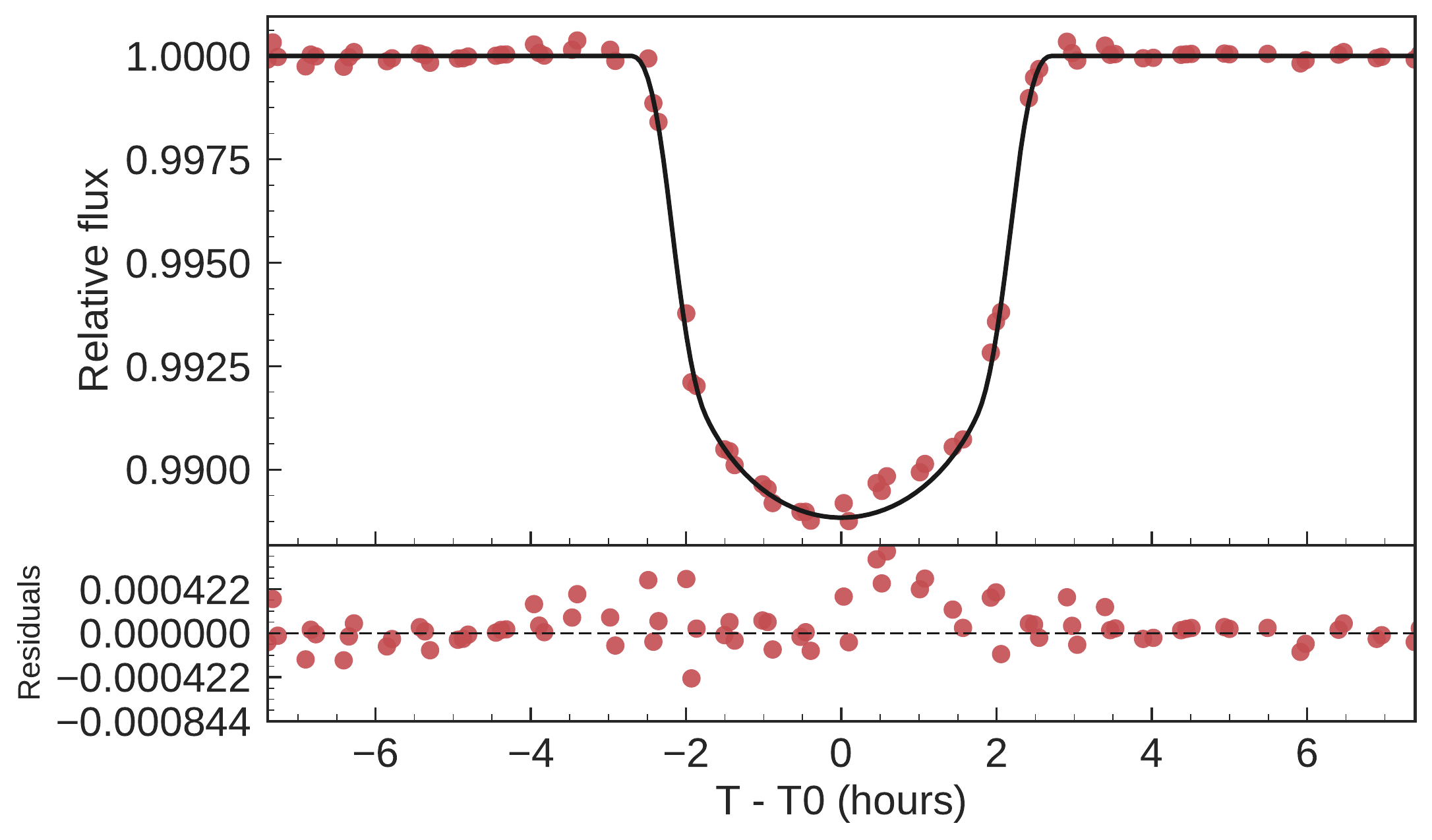}
\includegraphics[height=0.28\textwidth,trim={0 0.35cm 0 0},clip]{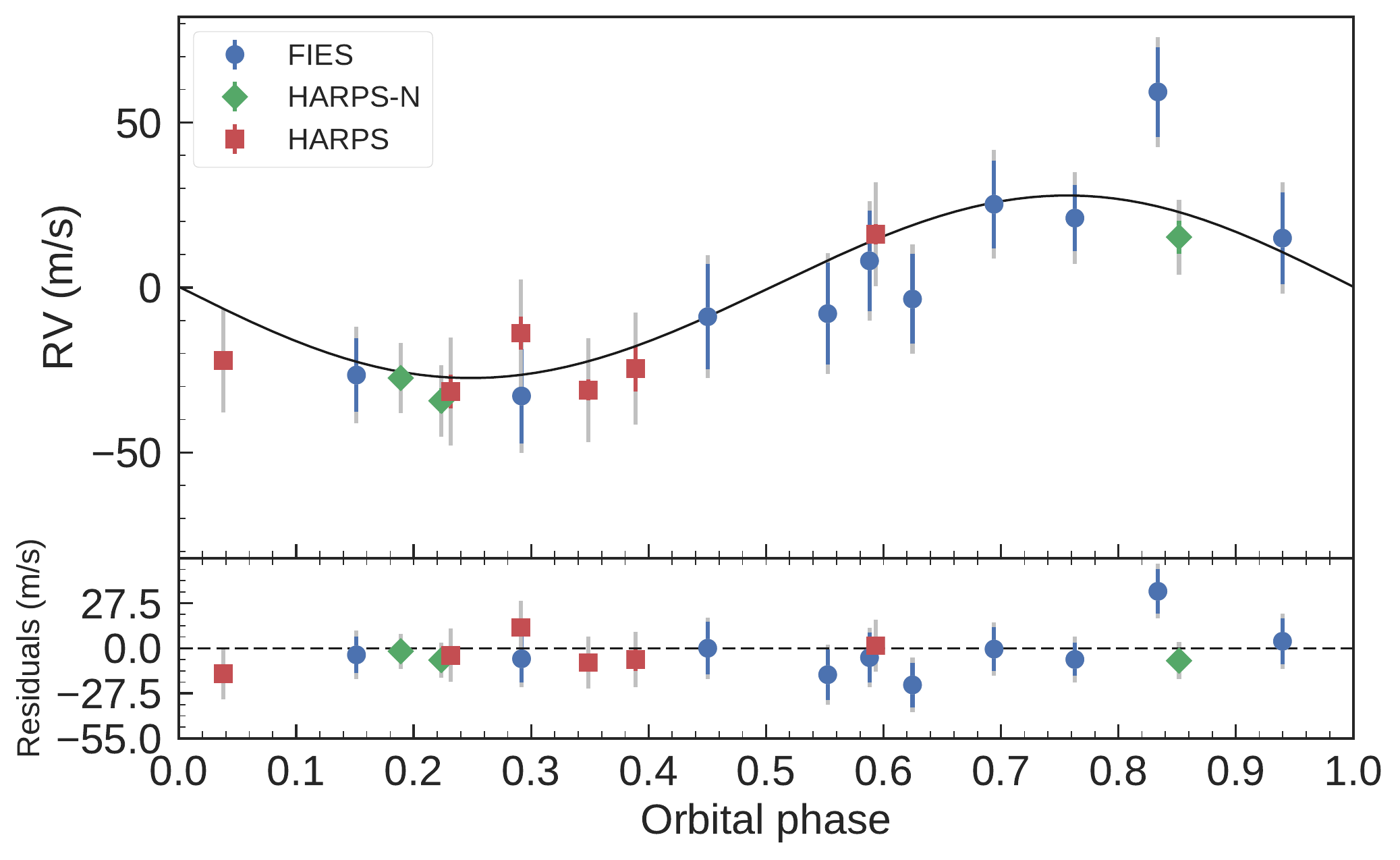} 
\caption{ \emph{Top}: FIES (blue circles), HARPS-N (green diamonds) and HARPS (red squares) RV measurements \emph{versus} time, following the subtraction of the systemic velocities for each instrument. The 1$\sigma$ uncertainties are marked using the same color used for each data-set. The vertical gray lines mark the error bars including jitter. The solid line represents the best fitting RV model, which includes the planet signal, and the activity signal at the stellar rotation period and its first harmonic. The dashed, dash-dotted, and dotted lines show the RV contribution of \pname, stellar rotation, and first harmonic, respectively. 
\emph{Lower left panel}: Transit light curve folded to the orbital period of \pname\ and residuals. The red points mark the \emph{K2} data and their error bars. The solid line mark the re-binned best-fitting transit model. \emph{ Lower right panel}: Phase-folded RV curve of \pname\ and best fitting Keplerian solution (solid line), following the subtraction of the two additional sinusoidal signals used to account for the stellar activity. The FIES, HARPS, and HARPS-N are corrected for the instrument offsets as derived from the global analysis. \label{fig:fits} }
\end{figure*}

To account for additional instrumental noise not included in the nominal RV error bars and/or imperfect treatment of the various sources of RV variations, we fitted for a jitter term for each instrument. The final parameter estimates and their error bars are listed in Table~\ref{tab:parameters}. They are defined as the median and the 68\% credible interval of the final posterior distributions. The best fitting transit and RV models are displayed in Figure~\ref{fig:fits} along with the photometric and RV data points.

\begin{table*}%[!th]
\begin{center}
\caption{Model comparison.\label{tab:models}}
\begin{tabular}{clcrccr}
\hline
\hline
\noalign{\smallskip}
Model & Comment & $\mathrm{N_{pars}}$ & $K_{\rm b}$ (m\,s$^{-1}$) & $\chi^2/\mathrm{dof}^{(a)}$ &  $\ln \mathcal{L}$ & AIC$^{(b)}$ \\
\hline
\noalign{\smallskip}
    \noalign{\smallskip}
P0 & Planet signal & 6 & $ 29.1 \pm 2.0 $ & 6.1 & 35.6 & -60 \\
    \noalign{\smallskip}
P1 & Planet signal + 1 sine-curve at $P_\mathrm{rot}$ & 9  & $ 29.4 \pm 2.4 $ & 3.4 & 58.1 & -98 \\ 
    \noalign{\smallskip}
P2 & Planet signal + 2 sine-curves at $P_\mathrm{rot}$ and $P_\mathrm{rot}$/2 & 11 & $27.3^{+2.6}_{-2.5}$  & 3.8 & 60.1 & -98 \\
    \noalign{\smallskip}
P3 & Planet signal + 3 sine-curves at $P_\mathrm{rot}$, $P_\mathrm{rot}$/2, and $P_\mathrm{rot}$/3 & 13 & $ 27.8_{- 2.6}^{+ 2.7} $ & 5.3 & 59.3 & -93 \\
    \noalign{\smallskip}
    NP1 & 1 sine-curve at $P_\mathrm{rot}$ (No planet signal) & 6 & $ 0 $ & 18.5 & -44.8 & 101 \\
    \noalign{\smallskip}
    NP2 & 2 sine-curves at $P_\mathrm{rot}$ and $P_\mathrm{rot}$/2 (No planet signal)  & 8 & $ 0 $ & 15.9 & -12.0 & 40 \\
\hline
\end{tabular}
  \begin{tablenotes}\footnotesize 
  \item \emph{Note} -- $^{(a)}$ $\chi^2$ value assuming no jitter. $^{(b)}$ We used the Akaike Information Criteria (AIC $=  2 \mathrm{N_{pars}} - \ln 2 \mathcal{L}$) instead of the widely used Bayesian information criteria (BIC) because our RV data sample is small (19 data points), and BIC performs better for large samples \citep{Burnham2002}.
\end{tablenotes}
\end{center}
\end{table*}

\subsection{Additional companion}

\cite{Huang2016} found that warm Jupiters with low eccentricities ($e \lesssim 0.4$) have inner low-mass companions. They used this evidence as an argument in favour of the \emph{in situ} formation, since the planet migration would have cleaned the warm Jupiter neighborhood. We searched the light curve for additional transit signals but found no evidence for an additional transiting planet in the system. As described in the previous paragraph, the periodogram of the RV residuals show no significant peak with false alarm probability lower than 5\,\%.

\subsection{Spot-crossing events}
\label{sec:spot_crossing}

The passage of a planet in front of a spot can be detected as a bump in the transit light curve \citep[see, e.g.,][]{2011ApJ...743...61S}. Spot-crossings events are clearly visible in the EVEREST transit light curves (Fig.~\ref{fig:fits}). The same features appear at the same times and with consistent amplitudes in the \citet{Vanderburg2014} data, confirming that the bumps are real and not due to systematics. To assess whether the bumps significantly affect the parameter estimates, we performed the joint analysis as described in Sect~\ref{sec:joint} including all the transit data points. We found that the final parameters are consistent within 1-$\sigma$ with those reported in Table~\ref{tab:parameters}.
 
\subsection{Planet's composition and formation scenario}

\begin{figure}
\includegraphics[width=0.5\textwidth]{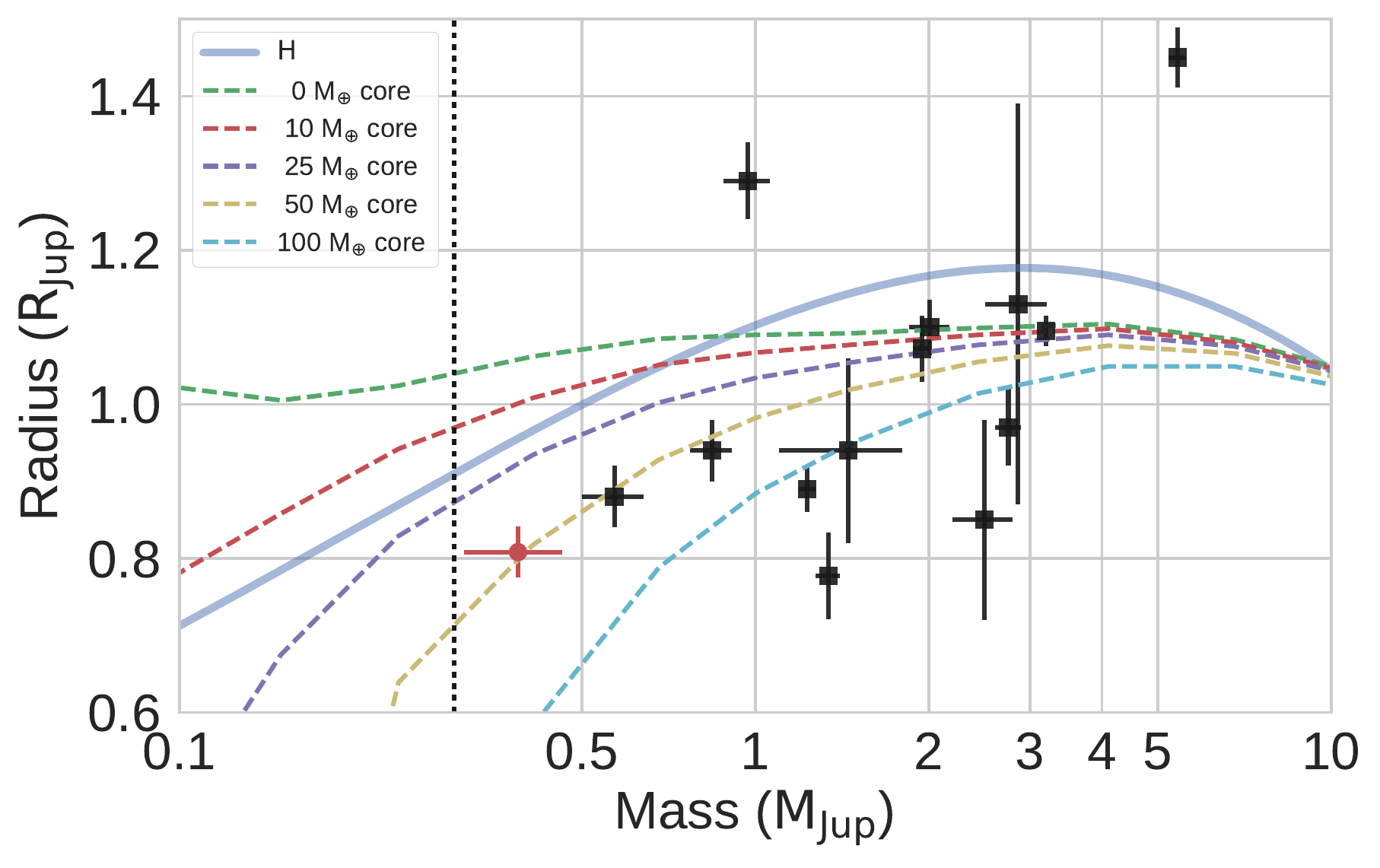}
\caption{Warm Jupiters (black squares; $M_\mathrm{p}$~$\geq$~0.3~$M_\mathrm{Jup}$ and 10~$\leq$~$P_\mathrm{orb}$~$\leq$~100~days) whose mass and radius have been estimated with a precision of at least 25\,\% (as of January 2017, exoplanet.eu). \pname\ is shown with a filled red circle. The solid line corresponds to a planet with a pure hydrogen composition \citep{2007ApJ...669.1279S}. The dashed lines represent the \citet{Fortney2007} models for planet core masses of 0, 10, 25, 50 and 100~$M_{\oplus}$. The vertical dotted line marks the giant planet lower limit as defined by \citet{Hatzes2015}. \label{fig:wj_plot}}
\end{figure}

With a mass of~$M_\mathrm{p}=$ \mpb\ and radius of $R_\mathrm{p}=$ \rpb\ (resulting in a mean density of $\rho_\mathrm{p}$=\denpb), \pname\ joins the small group of well characterized warm Jupiters. Fig.~\ref{fig:wj_plot} shows the position of \pname\ in the mass-radius diagram for warm Jupiters ($M_\mathrm{p}\,\geq\,0.3\,M_\mathrm{Jup}$; $10\,\leq P_\mathrm{orb}\,\leq\,100$\,days) whose mass and radius have been determined with a precision better than 25\,\% (14 objects). Notably, \pname\ is the transiting warm Jupiter with the lowest mass known to date, if the definition of giant planets given by \citet{Hatzes2015} is adopted. Fig.~\ref{fig:wj_plot} displays also the planetary models of \citet{Fortney2007} for different core masses and age between 1.0 and 4.5\,Gyrs. The planet radius of \pname\ can be explained if the planet has a core\footnote{Calculated by interpolating \citet{Fortney2007}'s models.} of \cmassp, containing $\sim$40\,\% of the total planetary mass. We expect that \pname\ has a solid core surrounded by a gaseous envelope.

\cite{Rafikov2006} found that a core of mass 5\,--\,20~$M_{\oplus}$ at a semi-major axis between 0.1 and 1.0\,AU would be able to start the runaway accretion phase to form a gas giant planet \emph{in situ}. However, according to his models, these kind of cores are unlikely to form, owing to the high irradiation coming from the star. \citet{Boley2016} suggested instead that more massive cores ($M_{\rm core} \gtrsim 20 M_{\oplus}$) can be built up from the merging of tightly packed inner planets formed at the early stages of the circumstellar disc. \citet{2016ApJ...829..114B} found a similar result and argued that the massive core of HD\,149026b ($M_{\rm core} \approx 100 M_{\oplus}$) could be explained by one or more super-Earths which merged and accreted the surrounding gas to form a gas-giant planet. \cite{Huang2016} suggested that these cores can initiate runaway accretion if they are formed in a region with enough gas around them, while those without enough volatiles remain super-Earths and represent the population of massive rocky planets unveiled by \emph{Kepler} around solar-like stars \citep[e.g.,][]{2014ApJ...789L..20D}. Based on these studies and given the semi-major axis of \ab, the $48\pm14\,M_\oplus$ core of \pname\ could have formed the planet \emph{in situ}. We note that the metallicity of \ssname\ is relatively high ([Fe/H] $= 0.21\pm0.05$), suggesting that the primordial circumstellar disc had a relatively high content of dust, which would have enhanced the formation of the core of \pname\ \citep[see, e.g.][]{2012ApJ...751...81J}. Alternatively, the planet might have formed beyond the snow line and migrated inwards via planet-disc interaction \citep[see, e.g.,][]{2014prpl.conf..667B}.

%---------------------------------------------------------------------------

\section{Conclusions}
\label{sec:conclusions}

We confirmed the planetary nature and derived the orbital and main physical parameters of \pname, a warm Jupiter ($T_\mathrm{eq}$\,=\,\Tequib) transiting an active (log\,$R^\prime_\mathrm{HK}=-4.46\pm0.06$) K0\,V star every 29 days. We measured a planetary mass of $M_\mathrm{p}$\,=\,\mpb\ and radius of $R_\mathrm{p}$\,=\,\rpb. At a separation of $a_\mathrm{p}$\,=\,\ab, the mean density of $\rho_\mathrm{p}$\,=\,\denpb\ implies that the planet has a core of \cmassp\  according to the evolutionary models of \citet{Fortney2007}. \pname\ joins the small group of well-characterized warm Jupiters whose mass and radius have been determined with a precision better than 25\,\%.

The spin-orbit angle, i.e., the angle between the spin axis of the star and the angular momentum vector of the orbit, can provide us with valuable information on the migration mechanisms of exoplanets \cite[see, e.g.][]{2010exop.book...55W, 2011ApJ...729..138M, 2012ApJ...757...18A, 2012A&A...543L...5G}. Currently, there are only 4 warm Jupiters ($M_\mathrm{p}$~$\geq$~0.3~$M_\mathrm{Jup}$ and 10~$\leq$~$P_\mathrm{orb}$~$\leq$~100~days) with measured obliquity\footnote{Source: \url{http://www2.mps.mpg.de/homes/heller/content/main_HRM.html}, as of January 2017.}. From this perspective, \ssname\ is an ideal target to measure the sky-project spin-orbit angle via observations of the Rossiter-McLaughlin (RM) effect. Assuming spin-orbit alignement, the expected amplitude of the RM anomaly is $\Delta \mathrm{RV}\,\approx\,\sqrt{1-b^2}\,(R_{\rm p}/R_{\star})^2\,$\vsini\,$\approx$\,25\,\ms\ \citep{2010exop.book...55W}. Given the brightness of the host star ($V$\,=\,11.653~mag), this amplitude can easily be measured using state-of-the-art spectrographs such as HARPS@ESO-3.6m. Moreover, the transit duration ($\sim$5 hours) is shorter than the visibility of \ssname, which is $\sim$9~hours  from La~Silla observatory (altitude higher than 30\degr\ above the horizon).

Alternatively, the spin-orbit angle could be measured from the analysis of the spot-crossing events as described in \citet{2011ApJ...733..127S} and \citet{2012Natur.487..449S}. Anomalies ascribable to the passage of \ssname b in front of stellar spots are visible in the 3 transit light curves observed by \emph{K2}. Unfortunately, the limited number of transits and the \emph{K2} long cadence data do not allow us to perform a meaningful quantitative analysis of the spot-crossing events. Given the amplitude of the detected anomalies ($\sim$0.1\,\%), space-based high-precision photometry is needed to detect the spot-crossing events. Observations performed with the upcoming CHaracterising ExOPlanets Satellite \citep[CHEOPS;][]{2013EPJWC..4703005B} would allow us to photometrically determine the spin-orbit angle of this system.

\section*{Acknowledgements}

We warmly thank the NOT, ESO, TNG staff members for their unique support during the observations. We are very thankful to Xavier Bonfils, Fran\c{c}ois Bouchy, Martin K\"urster, Tsevi Mazeh, Jorge Melendez, and Nuno Santos who kindly agreed to exchange HARPS and FIES time with us. Special thanks go to Antonino Lanza for assisting us with the calculation of the gyro-age~of~the~star. We also greatly thank the anonymous referee for her/his careful review and suggestions, which helped us to improve the manuscript. D.~Gandolfi gratefully acknowledges the financial support of the \emph{Programma Giovani Ricercatori -- Rita Levi Montalcini -- Rientro dei Cervelli (2012)} awarded by the Italian Ministry of Education, Universities and Research (MIUR). Sz. Csizmadia thanks the Hungarian OTKA Grant K113117. H.\,J. Deeg and D. Nespral acknowledge support by grant ESP2015-65712-C5-4-R of the Spanish Secretary of State for R\& D\&i (MINECO). D.\,Lorenzo-Oliveira acknowledges the support from FAPESP (2016/20667-8). This research was supported by the Ministerio de Economia y Competitividad under project FIS2012-31079. The research leading to these results has received funding from the European Union Seventh Framework Programme (FP7/2013-2016) under grant agreement No. 312430 (OPTICON). Based on observations obtained \emph{a}) with the Nordic Optical Telescope (NOT), operated on the island of La Palma jointly by Denmark, Finland, Iceland, Norway, and Sweden, in the Spanish Observatorio del Roque de los Muchachos (ORM) of the Instituto de Astrof\'isica de Canarias (IAC); \emph{b}) with the Italian Telescopio Nazionale Galileo (TNG) also operated at the ORM (IAC) on the island of La Palma by the INAF - Fundaci\'on Galileo Galilei; \emph{c}) the 3.6m ESO telescope at La Silla Observatory under programme ID 097.C-0948. The data presented here were obtained in part with ALFOSC, which is provided by the Instituto de Astrofisica de Andalucia (IAA) under a joint agreement with the University of Copenhagen and NOTSA. This paper includes data collected by the Kepler mission. Funding for the Kepler mission is provided by the NASA Science Mission directorate. Some of the data presented in this paper were obtained from the Mikulski Archive for Space Telescopes (MAST). STScI is operated by the Association of Universities for Research in Astronomy, Inc., under NASA contract NAS5-26555. Support for MAST for non-HST data is provided by the NASA Office of Space Science via grant NNX09AF08G and by other grants and contracts. M.F. and C.M.P. acknowledge generous support from the Swedish National Space Board. C. Eiroa and I. Rebollido are supported by Spanish grant  AYA2014-55840-P. P.D. acknowledge the support from INAF and Ministero dell'Istruzione, dell'Universit\`a e della Ricerca (MIUR) in the form of the grant ``Premiale VLT 2012'' and ``The Chemical and Dynamical Evolution of the Milky Way and Local Group Galaxies''. This work has made use of data from the European Space Agency (ESA) mission {\it Gaia} (\url{http://www.cosmos.esa.int/gaia}), processed by
the {\it Gaia} Data Processing and Analysis Consortium (DPAC, \url{http://www.cosmos.esa.int/web/gaia/dpac/consortium}). Funding for the DPAC has been provided by national institutions, in particular the institutions participating in the {\it Gaia} Multilateral Agreement.

%MNRAS table
\begin{table*}
  \caption{\ssname\ system parameters. \label{tab:parameters}}  
  \begin{tabular}{lcc}
  \hline
  Parameter & Prior$^{(\mathrm{a})}$ & Final value \\
  \hline
    %------------------------------------------------------
    \multicolumn{3}{l}{\emph{\bf{Stellar parameters}}} \\
    \noalign{\smallskip}
    %------------------------------------------------------
    Star mass $M_{\star}$ ($M_\odot$) &  $\cdots$ & \smass[] \\
    Star radius $R_{\star}$ ($R_\odot$) &  $\cdots$ & \sradius[]  \\
    Star density $\rho_\star$ (from spectroscopy, g\,cm$^{-3}$) & $\cdots$ & $2.02^{+0.25}_{-0.22}$  \\
    \noalign{\smallskip}    
    Star density $\rho_\star$ (from light curve, g\,cm$^{-3}$) & $\cdots$ & \densb[]  \\
	Effective Temperature $\mathrm{T_{eff}}$ (K) & $\cdots$ & $5340\pm110$ \\
    Surface gravity \logg\ (cgs) & $\cdots$ & $4.50\pm0.09$ \\ 
    Iron abundance [Fe/H] (dex) &  $\cdots$ & $0.22\pm0.08$ \\
    Microturbulent velocity \vmic\ (\kms) & $\cdots$& $0.9\pm0.1$ \\
    Macroturbulent velocity \vmac\ (\kms) &  $\cdots$ & $2.5\pm0.6$ \\
    Projected rotational velocity \vsini\ (\kms) &   $\cdots$ & $2.8\pm0.6$\\
    Rotational period $P_\mathrm{rot}$ (days) &  $\cdots$ & $17.24\pm0.12$ \\
    Activity index$^{(\mathrm{b})}$ log\,$R^\prime_\mathrm{HK}$  & $\cdots$ & $-4.46\pm0.06$ \\
    Gyrochronological age (Gyr) &  $\cdots$ & $1.8\pm0.3$ \\
    Interstellar extinction $A_\mathrm{V}$ (mag) &  $\cdots$ & $0.07\pm0.05$ \\
    Star distance $d$ (pc) & $\cdots$ & $152\pm10$ \\
  \hline
    %------------------------------------------------------
    \multicolumn{3}{l}{\emph{\bf{Model parameters of \pname}}} \\
    \noalign{\smallskip}
    %------------------------------------------------------
    Orbital period $P_{\mathrm{orb}}$ (days) &  $\mathcal{U}[ 28.3773 , 28.3873]$ & \Pb[] \\
    Transit epoch $T_0$ (BJD$_\mathrm{TDB}-$2\,450\,000) & $\mathcal{U}[7325.8120 , 7325.8220]$ & \Tzerob[]  \\ 
    Scaled semi-major axis $a/R_{\star}$ &  $\mathcal{U}[1.2,100]$ & \arb[] \\
    Planet-to-star radius ratio $R_\mathrm{p}/R_{\star}$ & $\mathcal{U}[0,0.2]$ & \rrb[]  \\
	Impact parameter, $b$  & $\mathcal{U}[0,1.2]$  & \bb[] \\
    $\sqrt{e} \sin \omega$ &  $\mathcal{U}[-1,1]$$^{(\mathrm{c})}$ & \esinb \\
    $\sqrt{e} \cos \omega$ &  $\mathcal{U}[-1,1]$$^{(\mathrm{c})}$ & \ecosb \\
    Radial velocity semi-amplitude variation $K$ (\ms) & $\mathcal{U}[0,200]$ & \kb[] \\
    \hline
    %------------------------------------------------------
    \multicolumn{3}{l}{\emph{\bf{Model parameters of RV sinusoidal signal at $P_\mathrm{rot}$}}} \\
    \noalign{\smallskip}
    %------------------------------------------------------
    Period $P_{\mathrm{rot}}$ (days) &  $\mathcal{N}[17.24,0.12]$ & \Pc[] \\
    Epoch $T_0$ (BJD$_\mathrm{TDB}-$2\,450\,000) & $\mathcal{U}[7324.0 , 7341.3]$ & \Tzeroc[]  \\  
    Radial velocity semi-amplitude variation $K$ (\ms) & $\mathcal{U}[0,200]$ & \kc[] \\
    \hline
   %------------------------------------------------------
    \multicolumn{3}{l}{\emph{\bf{Model parameters of RV sinusoidal signal at $P_\mathrm{rot}/2$}}} \\
    \noalign{\smallskip}
    %------------------------------------------------------
    Period $P_{\mathrm{orb}}$ (days) &  $\mathcal{F}[P_{\rm rot}/2]$ & \Pd[] \\
    Epoch $T_0$ (BJD$_\mathrm{TDB}-$2\,450\,000) & $\mathcal{U}[7317.0 , 7325.7]$ & \Tzerod[]  \\  
    Radial velocity semi-amplitude variation $K$ (m s$^{-1}$) & $\mathcal{U}[0,200]$ & \kd[] \\
    \hline
   %------------------------------------------------------
    \multicolumn{3}{l}{\emph{\bf{Additional model parameters}}} \\
    \noalign{\smallskip}
   %------------------------------------------------------
    Parameterized limb-darkening coefficient $q_1$  & $\mathcal{U}[0,1]$ & \qone \\
    \noalign{\smallskip}    
    Parameterized limb-darkening coefficient $q_2$  & $\mathcal{U}[0,1]$ & \qtwo \\
    Systemic velocity $\gamma_{\mathrm{FIES}}$  (km s$^{-1}$) & $\mathcal{U}[ -32.3913 , -30.2990]$ & \velFIES[]  \\      
    Systemic velocity $\gamma_{\mathrm{HARPS}}$  (km s$^{-1}$) & $\mathcal{U}[-32.2217 , -30.1633]$ & \velHARPS[] \\
    Systemic velocity $\gamma_{\mathrm{HARPS-N}}$  (km s$^{-1}$) & $\mathcal{U}[-32.2141 , -30.1683]$ & \velHARPSN[] \\
    \noalign{\smallskip}    
    Jitter term $\sigma_{\mathrm{FIES}}$  (m s$^{-1}$) & $\mathcal{U}[0,100]$ & \rvjitterFIES[]  \\
    Jitter term $\sigma_{\mathrm{HARPS}}$  (m s$^{-1}$) & $\mathcal{U}[0,100]$ & \rvjitterHARPS[] \\
    Jitter term $\sigma_{\mathrm{HARPS-N}}$  (m s$^{-1}$) & $\mathcal{U}[0,100]$ & \rvjitterHARPSN[] \\
    \hline 
   %------------------------------------------------------    
    \multicolumn{3}{l}{\emph{\bf{Derived parameters of \pname}}} \\
       \noalign{\smallskip}
   %------------------------------------------------------

    Planet mass $M_\mathrm{p}$ ($M_{\rm Jup}$) & $\cdots$ & \mpb[]  \\
    \noalign{\smallskip}
    Planet radius $R_\mathrm{p}$ ($R_{\rm Jup}$) & $\cdots$ & \rpb[] \\
    \noalign{\smallskip}
    Planet mean density $\rho_\mathrm{p}$ ($\mathrm{g\,cm^{-3}}$) & $\cdots$ & \denpb[] \\
    \noalign{\smallskip}    
    Semi-major axis of the planetary orbit $a$ (AU) & $\cdots$ & \ab[]  \\
    \noalign{\smallskip}
    Orbit eccentricity $e$ & $\cdots$ & \eb[]  \\
    \noalign{\smallskip}
    Argument of periastron of stellar orbit $\omega_{\star}$ (degrees) & $\cdots$ & \wb[] \\
    \noalign{\smallskip}
    Orbit inclination $i_\mathrm{p}$ (degrees) & $\cdots$ & \ib[] \\
    \noalign{\smallskip}
    Transit duration $\tau_{14}$ (hours) & $\cdots$ & \ttotb[] \\
    \noalign{\smallskip}
    Equilibrium temperature$^{(\mathrm{d})}$  $T_\mathrm{eq}$ (K)  & $\cdots$ &  \Tequib[] \\
  \hline
  \end{tabular}
  \begin{tablenotes}\footnotesize
  \item \emph{Note} -- The adopted Sun and Jupiter units follow the recommendations from the International Astronomical Union \citep{Prsa2016}. $^{(\mathrm{a})}$ $\mathcal{U}[a,b]$ refers to uniform priors between $a$ and $b$,  $\mathcal{N}[a,b]$ means Gaussian priors with  mean $a$ and standard deviation $b$ and $\mathcal{F}[a]$ to a fixed $a$ value. $^{(\mathrm{b})}$ Corrected for interstellar reddening following \citet{Fossati2017}. The correction is +0.06. $^{(\mathrm{c})}$ The code always ensures that $e < 1$. $^{(\mathrm{d})}$ Assuming albedo\,=\,0.
\end{tablenotes}
\end{table*}

\addcontentsline{toc}{section}{Acknowledgements}

%%%%%%%%%%%%%%%%%%%%%%%%%%%%%%%%%%%%%%%%%%%%%%%%%%



\bibliographystyle{mnras}
\bibliography{bibs} % if your bibtex file is called example.bib

\begin{thebibliography}{}
\makeatletter
\relax
\def\mn@urlcharsother{\let\do\@makeother \do\$\do\&\do\#\do\^\do\_\do\%\do\~}
\def\mn@doi{\begingroup\mn@urlcharsother \@ifnextchar [ {\mn@doi@}
  {\mn@doi@[]}}
\def\mn@doi@[#1]#2{\def\@tempa{#1}\ifx\@tempa\@empty \href
  {http://dx.doi.org/#2} {doi:#2}\else \href {http://dx.doi.org/#2} {#1}\fi
  \endgroup}
\def\mn@eprint#1#2{\mn@eprint@#1:#2::\@nil}
\def\mn@eprint@arXiv#1{\href {http://arxiv.org/abs/#1} {{\tt arXiv:#1}}}
\def\mn@eprint@dblp#1{\href {http://dblp.uni-trier.de/rec/bibtex/#1.xml}
  {dblp:#1}}
\def\mn@eprint@#1:#2:#3:#4\@nil{\def\@tempa {#1}\def\@tempb {#2}\def\@tempc
  {#3}\ifx \@tempc \@empty \let \@tempc \@tempb \let \@tempb \@tempa \fi \ifx
  \@tempb \@empty \def\@tempb {arXiv}\fi \@ifundefined
  {mn@eprint@\@tempb}{\@tempb:\@tempc}{\expandafter \expandafter \csname
  mn@eprint@\@tempb\endcsname \expandafter{\@tempc}}}

\bibitem[\protect\citeauthoryear{{Albrecht} et~al.,}{{Albrecht}
  et~al.}{2012}]{2012ApJ...757...18A}
{Albrecht} S.,  et~al., 2012, \mn@doi [\apj] {10.1088/0004-637X/757/1/18},
  \href {http://adsabs.harvard.edu/abs/2012ApJ...757...18A} {757, 18}

\bibitem[\protect\citeauthoryear{{Allard}, {Homeier}  \& {Freytag}}{{Allard}
  et~al.}{2011}]{Allard2011}
{Allard} F.,  {Homeier} D.,   {Freytag} B.,  2011, in {Johns-Krull} C.,
  {Browning} M.~K.,   {West} A.~A.,  eds,  Astronomical Society of the Pacific
  Conference Series Vol. 448, 16th Cambridge Workshop on Cool Stars, Stellar
  Systems, and the Sun. p.~91 (\mn@eprint {arXiv} {1011.5405})

\bibitem[\protect\citeauthoryear{{Antonini}, {Hamers}  \&
  {Lithwick}}{{Antonini} et~al.}{2016}]{Antonini2016}
{Antonini} F.,  {Hamers} A.~S.,   {Lithwick} Y.,  2016, preprint, \href
  {http://adsabs.harvard.edu/abs/2016arXiv160401781A} {} (\mn@eprint {arXiv}
  {1604.01781})

\bibitem[\protect\citeauthoryear{{Barnes}}{{Barnes}}{2010}]{Barnes2010b}
{Barnes} S.~A.,  2010, \mn@doi [\apj] {10.1088/0004-637X/722/1/222}, \href
  {http://adsabs.harvard.edu/abs/2010ApJ...722..222B} {722, 222}

\bibitem[\protect\citeauthoryear{{Barnes} \& {Kim}}{{Barnes} \&
  {Kim}}{2010}]{Barnes2010a}
{Barnes} S.~A.,  {Kim} Y.-C.,  2010, \mn@doi [\apj]
  {10.1088/0004-637X/721/1/675}, \href
  {http://adsabs.harvard.edu/abs/2010ApJ...721..675B} {721, 675}

\bibitem[\protect\citeauthoryear{{Barrag{\'a}n} \& {Gandolfi}}{{Barrag{\'a}n}
  \& {Gandolfi}}{2017}]{Barragan2017}
{Barrag{\'a}n} O.,  {Gandolfi} D.,  2017, {Exotrending}, Astrophysics Source
  Code Library (\mn@eprint {ascl} {1706.001})

\bibitem[\protect\citeauthoryear{{Barrag{\'a}n} et~al.,}{{Barrag{\'a}n}
  et~al.}{2016}]{Barragan2016}
{Barrag{\'a}n} O.,  et~al., 2016, \mn@doi [\aj] {10.3847/0004-6256/152/6/193},
  \href {http://adsabs.harvard.edu/abs/2016AJ....152..193B} {152, 193}

\bibitem[\protect\citeauthoryear{{Barrag{\'a}n}, {Gandolfi}  \&
  {Antoniciello}}{{Barrag{\'a}n} et~al.}{2017}]{pyaneti}
{Barrag{\'a}n} O.,  {Gandolfi} D.,   {Antoniciello} G.,  2017, {pyaneti},
  Astrophysics Source Code Library (\mn@eprint {ascl} {1707.003})

\bibitem[\protect\citeauthoryear{{Baruteau} et~al.,}{{Baruteau}
  et~al.}{2014}]{2014prpl.conf..667B}
{Baruteau} C.,  et~al., 2014, \mn@doi [Protostars and Planets VI]
  {10.2458/azu_uapress_9780816531240-ch029}, \href
  {http://adsabs.harvard.edu/abs/2014prpl.conf..667B} {pp 667--689}

\bibitem[\protect\citeauthoryear{{Batygin}, {Bodenheimer}  \&
  {Laughlin}}{{Batygin} et~al.}{2016}]{2016ApJ...829..114B}
{Batygin} K.,  {Bodenheimer} P.~H.,   {Laughlin} G.~P.,  2016, \mn@doi [\apj]
  {10.3847/0004-637X/829/2/114}, \href
  {http://adsabs.harvard.edu/abs/2016ApJ...829..114B} {829, 114}

\bibitem[\protect\citeauthoryear{{Boley}, {Granados Contreras}  \&
  {Gladman}}{{Boley} et~al.}{2016}]{Boley2016}
{Boley} A.~C.,  {Granados Contreras} A.~P.,   {Gladman} B.,  2016, \mn@doi
  [\apjl] {10.3847/2041-8205/817/2/L17}, \href
  {http://adsabs.harvard.edu/abs/2016ApJ...817L..17B} {817, L17}

\bibitem[\protect\citeauthoryear{{Boyajian} et~al.,}{{Boyajian}
  et~al.}{2013}]{Boyajian2013}
{Boyajian} T.~S.,  et~al., 2013, \mn@doi [\apj] {10.1088/0004-637X/771/1/40},
  \href {http://adsabs.harvard.edu/abs/2013ApJ...771...40B} {771, 40}

\bibitem[\protect\citeauthoryear{{Brahm} et~al.,}{{Brahm}
  et~al.}{2016}]{Brahm2016}
{Brahm} R.,  et~al., 2016, \mn@doi [\aj] {10.3847/0004-6256/151/4/89}, \href
  {http://adsabs.harvard.edu/abs/2016AJ....151...89B} {151, 89}

\bibitem[\protect\citeauthoryear{{Bressan}, {Marigo}, {Girardi}, {Salasnich},
  {Dal Cero}, {Rubele}  \& {Nanni}}{{Bressan} et~al.}{2012}]{Bressan2012}
{Bressan} A.,  {Marigo} P.,  {Girardi} L.,  {Salasnich} B.,  {Dal Cero} C.,
  {Rubele} S.,   {Nanni} A.,  2012, \mn@doi [\mnras]
  {10.1111/j.1365-2966.2012.21948.x}, \href
  {http://adsabs.harvard.edu/abs/2012MNRAS.427..127B} {427, 127}

\bibitem[\protect\citeauthoryear{{Broeg} et~al.,}{{Broeg}
  et~al.}{2013}]{2013EPJWC..4703005B}
{Broeg} C.,  et~al., 2013, in European Physical Journal Web of Conferences. p.
  03005 (\mn@eprint {arXiv} {1305.2270}), \mn@doi{10.1051/epjconf/20134703005}

\bibitem[\protect\citeauthoryear{{Bruntt} et~al.,}{{Bruntt}
  et~al.}{2010}]{Bruntt2010}
{Bruntt} H.,  et~al., 2010, \mn@doi [\mnras]
  {10.1111/j.1365-2966.2010.16575.x}, \href
  {http://adsabs.harvard.edu/abs/2010MNRAS.405.1907B} {405, 1907}

\bibitem[\protect\citeauthoryear{{Buchhave} et~al.,}{{Buchhave}
  et~al.}{2010}]{Buchhave2010}
{Buchhave} L.~A.,  et~al., 2010, \mn@doi [\apj] {10.1088/0004-637X/720/2/1118},
  \href {http://adsabs.harvard.edu/abs/2010ApJ...720.1118B} {720, 1118}

\bibitem[\protect\citeauthoryear{Burnham \& Anderson}{Burnham \&
  Anderson}{2002}]{Burnham2002}
Burnham K.,  Anderson D.,  2002, Model Selection and Multimodel Inference: A
  Practical Information-Theoretic Approach.
New York: Springer-Verlag

\bibitem[\protect\citeauthoryear{{Cabrera} et~al.,}{{Cabrera}
  et~al.}{2009}]{Cabrera2009}
{Cabrera} J.,  et~al., 2009, \mn@doi [\aap] {10.1051/0004-6361/200912684},
  \href {http://adsabs.harvard.edu/abs/2009A%26A...506..501C} {506, 501}

\bibitem[\protect\citeauthoryear{{Cabrera}, {Csizmadia}, {Erikson}, {Rauer}  \&
  {Kirste}}{{Cabrera} et~al.}{2012}]{Cabrera2012}
{Cabrera} J.,  {Csizmadia} S.,  {Erikson} A.,  {Rauer} H.,   {Kirste} S.,
  2012, \mn@doi [\aap] {10.1051/0004-6361/201219337}, \href
  {http://adsabs.harvard.edu/abs/2012A%26A...548A..44C} {548, A44}

\bibitem[\protect\citeauthoryear{{Cabrera} et~al.,}{{Cabrera}
  et~al.}{2014}]{Cabrera2014}
{Cabrera} J.,  et~al., 2014, \mn@doi [\apj] {10.1088/0004-637X/781/1/18}, \href
  {http://adsabs.harvard.edu/abs/2014ApJ...781...18C} {781, 18}

\bibitem[\protect\citeauthoryear{{Cantat-Gaudin} et~al.,}{{Cantat-Gaudin}
  et~al.}{2014}]{Cantat-Gaudin2014}
{Cantat-Gaudin} T.,  et~al., 2014, \mn@doi [\aap]
  {10.1051/0004-6361/201322533}, \href
  {http://adsabs.harvard.edu/abs/2014A%26A...562A..10C} {562, A10}

\bibitem[\protect\citeauthoryear{{Cardelli}, {Clayton}  \& {Mathis}}{{Cardelli}
  et~al.}{1989}]{Cardelli1989}
{Cardelli} J.~A.,  {Clayton} G.~C.,   {Mathis} J.~S.,  1989, \mn@doi [\apj]
  {10.1086/167900}, \href {http://adsabs.harvard.edu/abs/1989ApJ...345..245C}
  {345, 245}

\bibitem[\protect\citeauthoryear{{Carone} et~al.,}{{Carone}
  et~al.}{2012}]{Carone2012}
{Carone} L.,  et~al., 2012, \mn@doi [\aap] {10.1051/0004-6361/201116968}, \href
  {http://adsabs.harvard.edu/abs/2012A%26A...538A.112C} {538, A112}

\bibitem[\protect\citeauthoryear{{Carpano} et~al.,}{{Carpano}
  et~al.}{2009}]{Carpano2009}
{Carpano} S.,  et~al., 2009, \mn@doi [\aap] {10.1051/0004-6361/200911882},
  \href {http://adsabs.harvard.edu/abs/2009A%26A...506..491C} {506, 491}

\bibitem[\protect\citeauthoryear{{Castelli} \& {Kurucz}}{{Castelli} \&
  {Kurucz}}{2004}]{Castelli2004}
{Castelli} F.,  {Kurucz} R.~L.,  2004, preprint, \href
  {http://adsabs.harvard.edu/abs/2004astro.ph..5087C} {}

\bibitem[\protect\citeauthoryear{{Cavarroc} et~al.,}{{Cavarroc}
  et~al.}{2012}]{Cavarroc2012}
{Cavarroc} C.,  et~al., 2012, \mn@doi [\apss] {10.1007/s10509-011-0897-1},
  \href {http://adsabs.harvard.edu/abs/2012Ap%26SS.337..511C} {337, 511}

\bibitem[\protect\citeauthoryear{{Chabrier}}{{Chabrier}}{2001}]{Chabrier2001}
{Chabrier} G.,  2001, \mn@doi [\apj] {10.1086/321401}, \href
  {http://adsabs.harvard.edu/abs/2001ApJ...554.1274C} {554, 1274}

\bibitem[\protect\citeauthoryear{{Cosentino} et~al.,}{{Cosentino}
  et~al.}{2012}]{Cosentino2012}
{Cosentino} R.,  et~al., 2012, in Ground-based and Airborne Instrumentation for
  Astronomy IV. p. 84461V, \mn@doi{10.1117/12.925738}

\bibitem[\protect\citeauthoryear{{Cutri} et~al.,}{{Cutri}
  et~al.}{2003}]{Cutri2003}
{Cutri} R.~M.,  et~al., 2003, {2MASS All Sky Catalog of point sources.}

\bibitem[\protect\citeauthoryear{{Cutri} et~al.,}{{Cutri}
  et~al.}{2012}]{2012wise.rept....1C}
{Cutri} R.~M.,  et~al., 2012, Technical report, {Explanatory Supplement to the
  WISE All-Sky Data Release Products}

\bibitem[\protect\citeauthoryear{{Dawson}, {Johnson}, {Morton}, {Crepp},
  {Fabrycky}, {Murray-Clay}  \& {Howard}}{{Dawson}
  et~al.}{2012}]{2012ApJ...761..163D}
{Dawson} R.~I.,  {Johnson} J.~A.,  {Morton} T.~D.,  {Crepp} J.~R.,  {Fabrycky}
  D.~C.,  {Murray-Clay} R.~A.,   {Howard} A.~W.,  2012, \mn@doi [\apj]
  {10.1088/0004-637X/761/2/163}, \href
  {http://adsabs.harvard.edu/abs/2012ApJ...761..163D} {761, 163}

\bibitem[\protect\citeauthoryear{{Deeg} et~al.,}{{Deeg}
  et~al.}{2010}]{Deeg2010}
{Deeg} H.~J.,  et~al., 2010, \mn@doi [\nat] {10.1038/nature08856}, \href
  {http://adsabs.harvard.edu/abs/2010Natur.464..384D} {464, 384}

\bibitem[\protect\citeauthoryear{{Demory}}{{Demory}}{2014}]{2014ApJ...789L..20D}
{Demory} B.-O.,  2014, \mn@doi [\apjl] {10.1088/2041-8205/789/1/L20}, \href
  {http://adsabs.harvard.edu/abs/2014ApJ...789L..20D} {789, L20}

\bibitem[\protect\citeauthoryear{{Dong}, {Katz}  \& {Socrates}}{{Dong}
  et~al.}{2014}]{Dong2014a}
{Dong} S.,  {Katz} B.,   {Socrates} A.,  2014, \mn@doi [\apjl]
  {10.1088/2041-8205/781/1/L5}, \href
  {http://adsabs.harvard.edu/abs/2014ApJ...781L...5D} {781, L5}

\bibitem[\protect\citeauthoryear{{Doyle}, {Davies}, {Smalley}, {Chaplin}  \&
  {Elsworth}}{{Doyle} et~al.}{2014}]{Doyle2014}
{Doyle} A.~P.,  {Davies} G.~R.,  {Smalley} B.,  {Chaplin} W.~J.,   {Elsworth}
  Y.,  2014, \mn@doi [\mnras] {10.1093/mnras/stu1692}, \href
  {http://adsabs.harvard.edu/abs/2014MNRAS.444.3592D} {444, 3592}

\bibitem[\protect\citeauthoryear{{Dumusque}, {Boisse}  \& {Santos}}{{Dumusque}
  et~al.}{2014}]{2014ApJ...796..132D}
{Dumusque} X.,  {Boisse} I.,   {Santos} N.~C.,  2014, \mn@doi [\apj]
  {10.1088/0004-637X/796/2/132}, \href
  {http://adsabs.harvard.edu/abs/2014ApJ...796..132D} {796, 132}

\bibitem[\protect\citeauthoryear{{Endl} \& {Cochran}}{{Endl} \&
  {Cochran}}{2016}]{Endl2016}
{Endl} M.,  {Cochran} W.~D.,  2016, \mn@doi [\pasp]
  {10.1088/1538-3873/128/967/094502}, \href
  {http://adsabs.harvard.edu/abs/2016PASP..128i4502E} {128, 094502}

\bibitem[\protect\citeauthoryear{{Erikson} et~al.,}{{Erikson}
  et~al.}{2012}]{Erikson2012}
{Erikson} A.,  et~al., 2012, \mn@doi [\aap] {10.1051/0004-6361/201116934},
  \href {http://adsabs.harvard.edu/abs/2012A%26A...539A..14E} {539, A14}

\bibitem[\protect\citeauthoryear{{Fabricius} et~al.,}{{Fabricius}
  et~al.}{2016}]{Fabricius2016}
{Fabricius} C.,  et~al., 2016, \mn@doi [\aap] {10.1051/0004-6361/201628643},
  \href {http://adsabs.harvard.edu/abs/2016A%26A...595A...3F} {595, A3}

\bibitem[\protect\citeauthoryear{{Fortney}, {Marley}  \& {Barnes}}{{Fortney}
  et~al.}{2007}]{Fortney2007}
{Fortney} J.~J.,  {Marley} M.~S.,   {Barnes} J.~W.,  2007, \mn@doi [\apj]
  {10.1086/512120}, \href {http://adsabs.harvard.edu/abs/2007ApJ...659.1661F}
  {659, 1661}

\bibitem[\protect\citeauthoryear{{Fossati et al.}}{{Fossati et
  al.}}{2017}]{Fossati2017}
{Fossati et al.} 2017, A\&A, submitted

\bibitem[\protect\citeauthoryear{{Frandsen} \& {Lindberg}}{{Frandsen} \&
  {Lindberg}}{1999}]{Frandsen1999}
{Frandsen} S.,  {Lindberg} B.,  1999, in {Karttunen} H.,  {Piirola} V.,  eds,
  Astrophysics with the NOT. p.~71

\bibitem[\protect\citeauthoryear{{Frewen} \& {Hansen}}{{Frewen} \&
  {Hansen}}{2016}]{2016MNRAS.455.1538F}
{Frewen} S.~F.~N.,  {Hansen} B.~M.~S.,  2016, \mn@doi [\mnras]
  {10.1093/mnras/stv2322}, \href
  {http://adsabs.harvard.edu/abs/2016MNRAS.455.1538F} {455, 1538}

\bibitem[\protect\citeauthoryear{{Gandolfi} et~al.,}{{Gandolfi}
  et~al.}{2008}]{Gandolfi2008}
{Gandolfi} D.,  et~al., 2008, \mn@doi [\apj] {10.1086/591729}, \href
  {http://adsabs.harvard.edu/abs/2008ApJ...687.1303G} {687, 1303}

\bibitem[\protect\citeauthoryear{{Gandolfi} et~al.,}{{Gandolfi}
  et~al.}{2012}]{2012A&A...543L...5G}
{Gandolfi} D.,  et~al., 2012, \mn@doi [\aap] {10.1051/0004-6361/201219533},
  \href {http://adsabs.harvard.edu/abs/2012A%26A...543L...5G} {543, L5}

\bibitem[\protect\citeauthoryear{{Gandolfi} et~al.,}{{Gandolfi}
  et~al.}{2015}]{Gandolfi2015}
{Gandolfi} D.,  et~al., 2015, \mn@doi [\aap] {10.1051/0004-6361/201425062},
  \href {http://adsabs.harvard.edu/abs/2015A%26A...576A..11G} {576, A11}

\bibitem[\protect\citeauthoryear{{Gray}}{{Gray}}{1999}]{Gray1999}
{Gray} R.~O.,  1999, {SPECTRUM: A stellar spectral synthesis program},
  Astrophysics Source Code Library (\mn@eprint {ascl} {9910.002})

\bibitem[\protect\citeauthoryear{{Grziwa} \& {P{\"a}tzold}}{{Grziwa} \&
  {P{\"a}tzold}}{2016}]{Grziwa2016}
{Grziwa} S.,  {P{\"a}tzold} M.,  2016, preprint, \href
  {http://adsabs.harvard.edu/abs/2016arXiv160708417G} {} (\mn@eprint {arXiv}
  {1607.08417})

\bibitem[\protect\citeauthoryear{{Grziwa}, {P{\"a}tzold}  \& {Carone}}{{Grziwa}
  et~al.}{2012}]{Grziwa2012}
{Grziwa} S.,  {P{\"a}tzold} M.,   {Carone} L.,  2012, \mn@doi [\mnras]
  {10.1111/j.1365-2966.2011.19970.x}, \href
  {http://adsabs.harvard.edu/abs/2012MNRAS.420.1045G} {420, 1045}

\bibitem[\protect\citeauthoryear{{Hamers}, {Antonini}, {Lithwick}, {Perets}  \&
  {Portegies Zwart}}{{Hamers} et~al.}{2016}]{Hamers2016}
{Hamers} A.~S.,  {Antonini} F.,  {Lithwick} Y.,  {Perets} H.~B.,   {Portegies
  Zwart} S.~F.,  2016, preprint, \href
  {http://adsabs.harvard.edu/abs/2016arXiv160607438H} {} (\mn@eprint {arXiv}
  {1606.07438})

\bibitem[\protect\citeauthoryear{{Hatzes}}{{Hatzes}}{2002}]{Hatzes2002}
{Hatzes} A.~P.,  2002, \mn@doi [Astronomische Nachrichten]
  {10.1002/1521-3994(200208)323:3/4<392::AID-ASNA392>3.0.CO;2-M}, \href
  {http://adsabs.harvard.edu/abs/2002AN....323..392H} {323, 392}

\bibitem[\protect\citeauthoryear{{Hatzes} \& {Rauer}}{{Hatzes} \&
  {Rauer}}{2015}]{Hatzes2015}
{Hatzes} A.~P.,  {Rauer} H.,  2015, \mn@doi [\apjl]
  {10.1088/2041-8205/810/2/L25}, \href
  {http://adsabs.harvard.edu/abs/2015ApJ...810L..25H} {810, L25}

\bibitem[\protect\citeauthoryear{{Heiter} et~al.,}{{Heiter}
  et~al.}{2015}]{Heiter2015}
{Heiter} U.,  et~al., 2015, \mn@doi [\physscr] {10.1088/0031-8949/90/5/054010},
  \href {http://adsabs.harvard.edu/abs/2015PhyS...90e4010H} {90, 054010}

\bibitem[\protect\citeauthoryear{{Huang}, {Wu}  \& {Triaud}}{{Huang}
  et~al.}{2016}]{Huang2016}
{Huang} C.,  {Wu} Y.,   {Triaud} A.~H.~M.~J.,  2016, \mn@doi [\apj]
  {10.3847/0004-637X/825/2/98}, \href
  {http://adsabs.harvard.edu/abs/2016ApJ...825...98H} {825, 98}

\bibitem[\protect\citeauthoryear{{Jenkins} et~al.,}{{Jenkins}
  et~al.}{2017}]{Jenkins2017}
{Jenkins} J.~S.,  et~al., 2017, \mn@doi [\mnras] {10.1093/mnras/stw2811}, \href
  {http://adsabs.harvard.edu/abs/2017MNRAS.466..443J} {466, 443}

\bibitem[\protect\citeauthoryear{{Johnson} \& {Li}}{{Johnson} \&
  {Li}}{2012}]{2012ApJ...751...81J}
{Johnson} J.~L.,  {Li} H.,  2012, \mn@doi [\apj] {10.1088/0004-637X/751/2/81},
  \href {http://adsabs.harvard.edu/abs/2012ApJ...751...81J} {751, 81}

\bibitem[\protect\citeauthoryear{{Kipping}}{{Kipping}}{2010}]{Kipping2010}
{Kipping} D.~M.,  2010, \mn@doi [\mnras] {10.1111/j.1365-2966.2010.17242.x},
  \href {http://adsabs.harvard.edu/abs/2010MNRAS.408.1758K} {408, 1758}

\bibitem[\protect\citeauthoryear{{Kley} \& {Nelson}}{{Kley} \&
  {Nelson}}{2012}]{2012ARA&A..50..211K}
{Kley} W.,  {Nelson} R.~P.,  2012, \mn@doi [\araa]
  {10.1146/annurev-astro-081811-125523}, \href
  {http://adsabs.harvard.edu/abs/2012ARA%26A..50..211K} {50, 211}

\bibitem[\protect\citeauthoryear{{Kov{\'a}cs}, {Zucker}  \&
  {Mazeh}}{{Kov{\'a}cs} et~al.}{2002}]{Kovacs2002}
{Kov{\'a}cs} G.,  {Zucker} S.,   {Mazeh} T.,  2002, \mn@doi [\aap]
  {10.1051/0004-6361:20020802}, \href
  {http://adsabs.harvard.edu/abs/2002A%26A...391..369K} {391, 369}

\bibitem[\protect\citeauthoryear{{Kuerster}, {Schmitt}, {Cutispoto}  \&
  {Dennerl}}{{Kuerster} et~al.}{1997}]{Kuerster1997}
{Kuerster} M.,  {Schmitt} J.~H.~M.~M.,  {Cutispoto} G.,   {Dennerl} K.,  1997,
  \aap, \href {http://adsabs.harvard.edu/abs/1997A%26A...320..831K} {320, 831}

\bibitem[\protect\citeauthoryear{{Kurucz}}{{Kurucz}}{2013}]{Kurucz2013}
{Kurucz} R.~L.,  2013, {ATLAS12: Opacity sampling model atmosphere program},
  Astrophysics Source Code Library (\mn@eprint {ascl} {1303.024})

\bibitem[\protect\citeauthoryear{{Lenz} \& {Breger}}{{Lenz} \&
  {Breger}}{2005}]{Lenz2005}
{Lenz} P.,  {Breger} M.,  2005, \mn@doi [Communications in Asteroseismology]
  {10.1553/cia146s53}, \href
  {http://adsabs.harvard.edu/abs/2005CoAst.146...53L} {146, 53}

\bibitem[\protect\citeauthoryear{{Luger}, {Agol}, {Kruse}, {Barnes}, {Becker},
  {Foreman-Mackey}  \& {Deming}}{{Luger} et~al.}{2016}]{2016AJ....152..100L}
{Luger} R.,  {Agol} E.,  {Kruse} E.,  {Barnes} R.,  {Becker} A.,
  {Foreman-Mackey} D.,   {Deming} D.,  2016, \mn@doi [\aj]
  {10.3847/0004-6256/152/4/100}, \href
  {http://adsabs.harvard.edu/abs/2016AJ....152..100L} {152, 100}

\bibitem[\protect\citeauthoryear{{Magrini} et~al.,}{{Magrini}
  et~al.}{2013}]{Magrini2013}
{Magrini} L.,  et~al., 2013, \mn@doi [\aap] {10.1051/0004-6361/201321844},
  \href {http://adsabs.harvard.edu/abs/2013A%26A...558A..38M} {558, A38}

\bibitem[\protect\citeauthoryear{{Mandel} \& {Agol}}{{Mandel} \&
  {Agol}}{2002}]{2002ApJ...580L.171M}
{Mandel} K.,  {Agol} E.,  2002, \mn@doi [\apjl] {10.1086/345520}, \href
  {http://adsabs.harvard.edu/abs/2002ApJ...580L.171M} {580, L171}

\bibitem[\protect\citeauthoryear{{Mayor} et~al.,}{{Mayor}
  et~al.}{2003}]{Mayor2003}
{Mayor} M.,  et~al., 2003, The Messenger, \href
  {http://adsabs.harvard.edu/abs/2003Msngr.114...20M} {114, 20}

\bibitem[\protect\citeauthoryear{{McQuillan}, {Mazeh}  \&
  {Aigrain}}{{McQuillan} et~al.}{2014}]{2014ApJS..211...24M}
{McQuillan} A.,  {Mazeh} T.,   {Aigrain} S.,  2014, \mn@doi [\apjs]
  {10.1088/0067-0049/211/2/24}, \href
  {http://adsabs.harvard.edu/abs/2014ApJS..211...24M} {211, 24}

\bibitem[\protect\citeauthoryear{{Morton} \& {Johnson}}{{Morton} \&
  {Johnson}}{2011}]{2011ApJ...729..138M}
{Morton} T.~D.,  {Johnson} J.~A.,  2011, \mn@doi [\apj]
  {10.1088/0004-637X/729/2/138}, \href
  {http://adsabs.harvard.edu/abs/2011ApJ...729..138M} {729, 138}

\bibitem[\protect\citeauthoryear{{Niedzielski} et~al.,}{{Niedzielski}
  et~al.}{2016}]{Niedzielski2016}
{Niedzielski} A.,  et~al., 2016, preprint, \href
  {http://adsabs.harvard.edu/abs/2016arXiv160307581N} {} (\mn@eprint {arXiv}
  {1603.07581})

\bibitem[\protect\citeauthoryear{{Ortiz} et~al.,}{{Ortiz}
  et~al.}{2015}]{Ortiz2014}
{Ortiz} M.,  et~al., 2015, \mn@doi [\aap] {10.1051/0004-6361/201425146}, \href
  {http://adsabs.harvard.edu/abs/2015A%26A...573L...6O} {573, L6}

\bibitem[\protect\citeauthoryear{{Pepe} et~al.,}{{Pepe}
  et~al.}{2013}]{Pepe2013}
{Pepe} F.,  et~al., 2013, \mn@doi [\nat] {10.1038/nature12768}, \href
  {http://adsabs.harvard.edu/abs/2013Natur.503..377P} {503, 377}

\bibitem[\protect\citeauthoryear{{Petrovich} \& {Tremaine}}{{Petrovich} \&
  {Tremaine}}{2016}]{Petrovich2016}
{Petrovich} C.,  {Tremaine} S.,  2016, preprint, \href
  {http://adsabs.harvard.edu/abs/2016arXiv160400010P} {} (\mn@eprint {arXiv}
  {1604.00010})

\bibitem[\protect\citeauthoryear{{Pr{\v s}a} et~al.,}{{Pr{\v s}a}
  et~al.}{2016}]{Prsa2016}
{Pr{\v s}a} A.,  et~al., 2016, \mn@doi [\aj] {10.3847/0004-6256/152/2/41},
  \href {http://adsabs.harvard.edu/abs/2016AJ....152...41P} {152, 41}

\bibitem[\protect\citeauthoryear{{Rafikov}}{{Rafikov}}{2006}]{Rafikov2006}
{Rafikov} R.~R.,  2006, \mn@doi [\apj] {10.1086/505695}, \href
  {http://adsabs.harvard.edu/abs/2006ApJ...648..666R} {648, 666}

\bibitem[\protect\citeauthoryear{{Ryabchikova}, {Pakhomov}  \&
  {Piskunov}}{{Ryabchikova} et~al.}{2011}]{Ryabchikova2011}
{Ryabchikova} T.~A.,  {Pakhomov} Y.~V.,   {Piskunov} N.~E.,  2011, Kazan
  Izdatel Kazanskogo Universiteta, \href
  {http://adsabs.harvard.edu/abs/2011KIzKU.153...61R} {153, 61}

\bibitem[\protect\citeauthoryear{{Saad-Olivera}, {Nesvorn{\'y}}, {Kipping}  \&
  {Roig}}{{Saad-Olivera} et~al.}{2017}]{Saad-Olivera2017}
{Saad-Olivera} X.,  {Nesvorn{\'y}} D.,  {Kipping} D.~M.,   {Roig} F.,  2017,
  \mn@doi [\aj] {10.3847/1538-3881/aa64e0}, \href
  {http://adsabs.harvard.edu/abs/2017AJ....153..198S} {153, 198}

\bibitem[\protect\citeauthoryear{{Saar} \& {Donahue}}{{Saar} \&
  {Donahue}}{1997}]{Saar1997}
{Saar} S.~H.,  {Donahue} R.~A.,  1997, \mn@doi [\apj] {10.1086/304392}, \href
  {http://adsabs.harvard.edu/abs/1997ApJ...485..319S} {485, 319}

\bibitem[\protect\citeauthoryear{{Sanchis-Ojeda} \& {Winn}}{{Sanchis-Ojeda} \&
  {Winn}}{2011}]{2011ApJ...743...61S}
{Sanchis-Ojeda} R.,  {Winn} J.~N.,  2011, \mn@doi [\apj]
  {10.1088/0004-637X/743/1/61}, \href
  {http://adsabs.harvard.edu/abs/2011ApJ...743...61S} {743, 61}

\bibitem[\protect\citeauthoryear{{Sanchis-Ojeda}, {Winn}, {Holman}, {Carter},
  {Osip}  \& {Fuentes}}{{Sanchis-Ojeda} et~al.}{2011}]{2011ApJ...733..127S}
{Sanchis-Ojeda} R.,  {Winn} J.~N.,  {Holman} M.~J.,  {Carter} J.~A.,  {Osip}
  D.~J.,   {Fuentes} C.~I.,  2011, \mn@doi [\apj]
  {10.1088/0004-637X/733/2/127}, \href
  {http://adsabs.harvard.edu/abs/2011ApJ...733..127S} {733, 127}

\bibitem[\protect\citeauthoryear{{Sanchis-Ojeda} et~al.,}{{Sanchis-Ojeda}
  et~al.}{2012}]{2012Natur.487..449S}
{Sanchis-Ojeda} R.,  et~al., 2012, \mn@doi [\nat] {10.1038/nature11301}, \href
  {http://adsabs.harvard.edu/abs/2012Natur.487..449S} {487, 449}

\bibitem[\protect\citeauthoryear{{Schlegel}, {Finkbeiner}  \&
  {Davis}}{{Schlegel} et~al.}{1998}]{Schlegel1998}
{Schlegel} D.~J.,  {Finkbeiner} D.~P.,   {Davis} M.,  1998, \mn@doi [\apj]
  {10.1086/305772}, \href {http://adsabs.harvard.edu/abs/1998ApJ...500..525S}
  {500, 525}

\bibitem[\protect\citeauthoryear{{Seager}, {Kuchner}, {Hier-Majumder}  \&
  {Militzer}}{{Seager} et~al.}{2007}]{2007ApJ...669.1279S}
{Seager} S.,  {Kuchner} M.,  {Hier-Majumder} C.~A.,   {Militzer} B.,  2007,
  \mn@doi [\apj] {10.1086/521346}, \href
  {http://adsabs.harvard.edu/abs/2007ApJ...669.1279S} {669, 1279}

\bibitem[\protect\citeauthoryear{{Smith} et~al.,}{{Smith}
  et~al.}{2017}]{2017MNRAS.464.2708S}
{Smith} A.~M.~S.,  et~al., 2017, \mn@doi [\mnras] {10.1093/mnras/stw2487},
  \href {http://adsabs.harvard.edu/abs/2017MNRAS.464.2708S} {464, 2708}

\bibitem[\protect\citeauthoryear{{Sneden}, {Bean}, {Ivans}, {Lucatello}  \&
  {Sobeck}}{{Sneden} et~al.}{2012}]{Sneden2012}
{Sneden} C.,  {Bean} J.,  {Ivans} I.,  {Lucatello} S.,   {Sobeck} J.,  2012,
  {MOOG: LTE line analysis and spectrum synthesis}, Astrophysics Source Code
  Library (\mn@eprint {ascl} {1202.009})

\bibitem[\protect\citeauthoryear{{Stetson} \& {Pancino}}{{Stetson} \&
  {Pancino}}{2008}]{Stetson2008}
{Stetson} P.~B.,  {Pancino} E.,  2008, \mn@doi [\pasp] {10.1086/596126}, \href
  {http://adsabs.harvard.edu/abs/2008PASP..120.1332S} {120, 1332}

\bibitem[\protect\citeauthoryear{{Telting} et~al.,}{{Telting}
  et~al.}{2014}]{Telting2014}
{Telting} J.~H.,  et~al., 2014, \mn@doi [Astronomische Nachrichten]
  {10.1002/asna.201312007}, \href
  {http://adsabs.harvard.edu/abs/2014AN....335...41T} {335, 41}

\bibitem[\protect\citeauthoryear{{Tull}, {MacQueen}, {Sneden}  \&
  {Lambert}}{{Tull} et~al.}{1995}]{Tull1995}
{Tull} R.~G.,  {MacQueen} P.~J.,  {Sneden} C.,   {Lambert} D.~L.,  1995,
  \mn@doi [\pasp] {10.1086/133548}, \href
  {http://adsabs.harvard.edu/abs/1995PASP..107..251T} {107, 251}

\bibitem[\protect\citeauthoryear{{Valenti} \& {Fischer}}{{Valenti} \&
  {Fischer}}{2005}]{vf05}
{Valenti} J.~A.,  {Fischer} D.~A.,  2005, \mn@doi [\apjs] {10.1086/430500},
  \href {http://adsabs.harvard.edu/abs/2005ApJS..159..141V} {159, 141}

\bibitem[\protect\citeauthoryear{{Valenti} \& {Piskunov}}{{Valenti} \&
  {Piskunov}}{1996}]{vp96}
{Valenti} J.~A.,  {Piskunov} N.,  1996, \aaps, \href
  {http://adsabs.harvard.edu/abs/1996A%26AS..118..595V} {118, 595}

\bibitem[\protect\citeauthoryear{{Vanderburg} \& {Johnson}}{{Vanderburg} \&
  {Johnson}}{2014}]{Vanderburg2014}
{Vanderburg} A.,  {Johnson} J.~A.,  2014, \mn@doi [\pasp] {10.1086/678764},
  \href {http://adsabs.harvard.edu/abs/2014PASP..126..948V} {126, 948}

\bibitem[\protect\citeauthoryear{{Winn}}{{Winn}}{2010}]{2010exop.book...55W}
{Winn} J.~N.,  2010, {Exoplanet Transits and Occultations}.
University of Arizona Press, pp 55--77

\bibitem[\protect\citeauthoryear{{Zechmeister} \& {K{\"u}rster}}{{Zechmeister}
  \& {K{\"u}rster}}{2009}]{2009A&A...496..577Z}
{Zechmeister} M.,  {K{\"u}rster} M.,  2009, \mn@doi [\aap]
  {10.1051/0004-6361:200811296}, \href
  {http://adsabs.harvard.edu/abs/2009A%26A...496..577Z} {496, 577}

\bibitem[\protect\citeauthoryear{{da Silva} et~al.,}{{da Silva}
  et~al.}{2007}]{daSilva2007}
{da Silva} R.,  et~al., 2007, \mn@doi [\aap] {10.1051/0004-6361:20077314},
  \href {http://adsabs.harvard.edu/abs/2007A%26A...473..323D} {473, 323}

\makeatother
\end{thebibliography}

%\begin{thebibliography}{99}

%\end{thebibliography}

%%%%%%%%%%%%%%%%%%%%%%%%%%%%%%%%%%%%%%%%%%%%%%%%%%

%%%%%%%%%%%%%%%%% APPENDICES %%%%%%%%%%%%%%%%%%%%%

%%%%%%%%%%%%%%%%%%%%%%%%%%%%%%%%%%%%%%%%%%%%%%%%%%

% Don't change these lines
\bsp	% typesetting comment
\label{lastpage}
\end{document}